\documentclass[preprint,12pt,authoryear]{elsarticle}

\usepackage{amssymb}
\usepackage{amsmath}
\usepackage{algorithm}
\usepackage{algpseudocode}
\usepackage{graphicx}
\usepackage{subcaption}
\usepackage{geometry}
\usepackage{lipsum} 
\usepackage{float} 

\journal{Neurocomputing}

\begin{document}
\begin{frontmatter}
\title{Integrated Automated Car Following and Lane-changing control based on a Parametrized Deep Q-network with Hybrid Action Space} 


\author[label1]{Hao Zhang} 
\author[label1]{Zihao Li} 
\author[label1]{Yang Zhou} 
\affiliation[label1]{
            addressline={Zachry Department of Civil $\&$ Environmental Engineering, Texas A$\&$M University, College Station, TX 77843, USA}, 
            }

\begin{abstract}
Lane-change, a triggering of traffic disturbances to the upstream vehicles, is detrimental to traffic safety and efficiency. Coupled with car following behavior, the joint maneuvers depict the general picture of how traffic disturbances generate and propagate through vehicle streams, especially under traffic congestion. This study proposes an integrated control framework for lane-changing and car-following for connected and automated vehicles (CAVs), where those two tasks are largely treated as independent driving tasks by prevailing methods. Utilizing the Parametrized Deep Q-Network (P-DQN) with a hybrid action space, the framework adeptly models multiple objectives in CAV control. The P-DQN’s high-level control is employed for discrete lane change decisions, while its low-level control manages continuous acceleration actions (i.e., lateral and longitudinal). These actions are interdependently determined, seamlessly integrating car-following and lane-changing control. By training to maximize cumulative rewards, the proposed control strategy ensures driving safety as well as the efficiency of car-following, lane-changing, and lane-keeping. Through numerical experiments, it is indicated that the P-DQN outperforms separated control methods (e.g., the combination of Minimizing Overall Braking Decelerations Induced by Lane Changes (MOBIL) model and the Intelligent Driver Model (IDM)) in terms of safety and comfort.
\end{abstract}


\end{frontmatter}



\section{Introduction}
\label{sec1}

The advent of autonomous driving technology has brought forth a myriad of opportunities for enhancing road safety and traffic efficiency \citep{li2024disturbances}. Central to this technological evolution is the need for sophisticated control strategies capable of making reliable decision in complex driving scenarios. Lane changing behavior, a critical aspect of automated vehicle control due to its complexity and the need for precise decision making in dynamic traffic condition, directly impacts safety and traffic flow efficiency \citep{elallid2022comprehensive,li2022decision,nian2020review,zhang2023identifying,zhang2023spatiotemporal}. The traditional approaches to model lane-changing behaviors mainly involve pre-defined rules and explicitly designed models, such as Minimizing Overall Braking decelerations Induced by Lane changes (MOBIL) \citep{wang2015game}. MOBIL utilizes a utility function to determine whether a vehicle should change lanes. The function examines the benefits and potential negative impacts of a lane change, focusing on its effects on both the vehicle initiating the change and the surrounding traffic. Additionally, several studies have proposed the concept of a virtual lane-change trajectory, which vehicle can follow during the lane changing process \citep{choi2015lane,ho2009lane}. They usually utilize mathematical functions such as quintic polynomial, sinusoidal, and trapezoidal to approximate lane-change trajectories  \citep{zhou2017lane}. While these models perform reasonably well in designed scenarios or within the model’s boundaries, they are insufficient for handling situations beyond the defined range \citep{ammourah2023deep}. For example, MOBIL is incapable of planning for a further lane-change maneuver \citep{treiber2016mobil}. These models cannot fully describe the lane-changing behavior, because the lane-changing decision is related to multiple factors, including safety considerations, driver comfort, and the efficiency of traffic flow. Relying solely on predefined rules and explicit models in connected automated vehicles (CAVs) control may fall short in effectively managing the dynamic and stochastic nature of traffic scenarios. To address these limitations, instead of precisely modeling lane-changing behavior, the focus is shifting towards directly employing data-driven approaches for controlling CAVs under lane-changing, an area that is increasingly attracting the interest of researchers.

Reinforcement Learning (RL)-based approaches emerge as a promising method because they enable CAVs to learn suitable behaviors through trial and error within a variety of traffic conditions. Besides, RL-based approach can integrate various factors, such as safety constraints, driver comfort, and traffic efficiency, into their reward functions. This integration ensures that the learned policy aims to achieve the maximum designed rewards, providing a more comprehensive and adaptable solution compared to model-based approaches \citep{shi2022deep,shi2023physics}. Thus, several studies have utilized RL to control the AVs during lane-changing process \citep{aradi2020survey,huegle2019dynamic,zhu2021survey}. Because lane-changing control usually involves multiple subtasks, including adjusting speed to utilize an acceptable gap, lane change decision making, and lane change implementation \citep{xie2019data}, existing RL-based lane-changing control treat each subtask individually and propose separate control strategies \citep{you2019autonomous,li2022combining}. For instance, \cite{wang2018reinforcement} implemented the Intelligent Driver Model (IDM) for longitudinal control and a gap-selection model to ensure a safe lane-changing gap, while using RL for executing smooth cut-in maneuvers.\cite{ye2020automated} introduced a proximal policy optimization-based RL to make an upper-level decision (e.g., lane keeping, lane change decision making). In the lower-level control, IDM and the lane change model built in SUMO are utilized to execute longitudinal and lateral control.

\begin{figure}[htbp]
    \centering
    \includegraphics[width=0.6\linewidth]{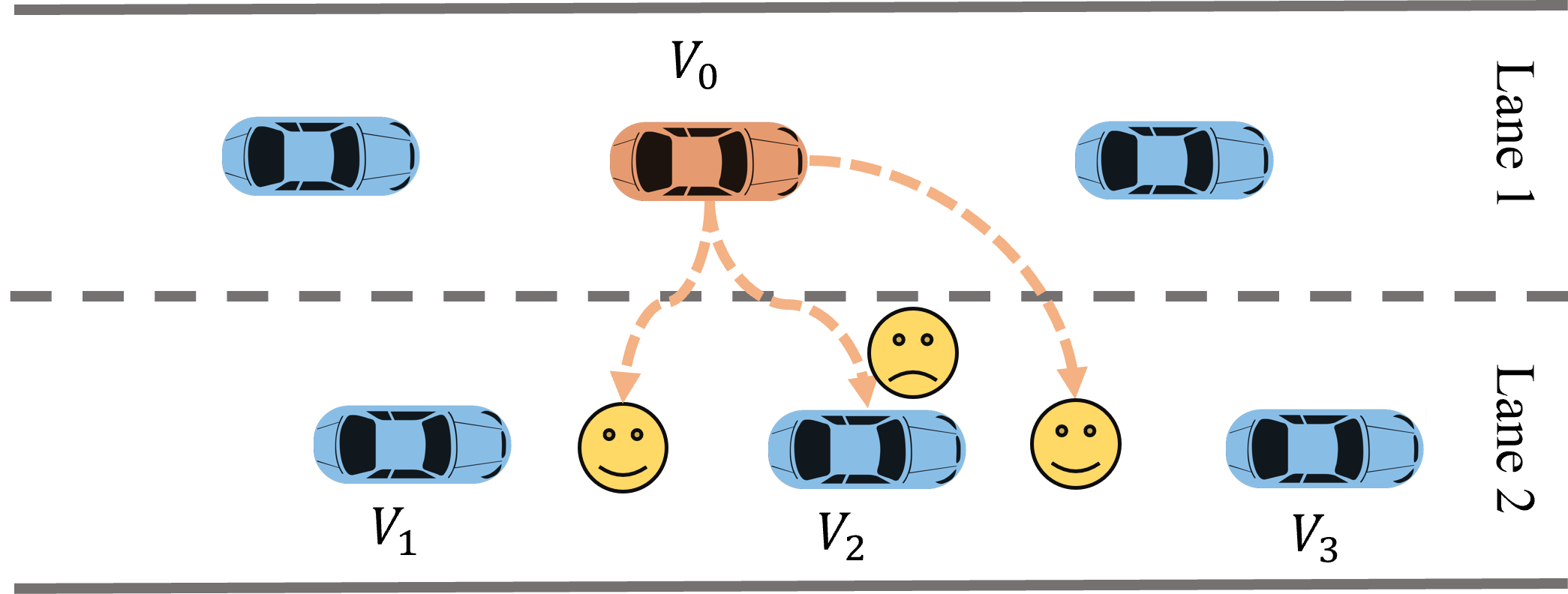}
    \caption{Illustration of lane changing maneuvers}
    \label{fig:1}
\end{figure}

Although these studies circumvent the need to implement multiple subtasks, they nevertheless overlook the interactions among these subtasks. The lane-changing maneuver should be treated as an integrated movement, rather than treating as several independent subtasks \citep{zhang2019simultaneous,toledo2007integrated,huang2020data}. In other words, car-following and lane-changing are interconnected and mutually influenced processes. As an example in Fig. \ref{fig:1}, ego vehicle $V_0$ wants to change to lane 2 and both the gaps between $V_1$ and $V_2$, and between $V_2$ and $V_3$ are acceptable for the ego vehicle’s lane change. Based on the MOBIL model, $V_0$ will reject to change to the lane 2 due to a potential collision risk with $V_2$. If the interaction between lane-changing and car-following is not considered and $V_0$ and $V_2$ have the same speed, $V_0$ will always reject the lane changing until it becomes urgent because $V_0$ will not change its speed to utilize the acceptable gaps. This situation is different from realistic driving behaviors. $V_0$ can perform lane change using the gap between $V_2$ and $V_3$ by accelerating or using the gap between $V_1$ and $V_2$ by decelerating. However, the integrated control of lane-changing and car-following is challenge in RL because its action space is hybrid \citep{peng2022integrated}.Specially, determining when and where to change lanes constitutes a discrete action, while the lane change implementation and car-following involve continuous actions. Traditional RL-based methods struggle to address the challenge of hybrid action space. \cite{hoel2018automated} and \cite{zhang2019discretionary} have attempted to achieve this integrated control in RL by modifying the action space. Commonly, discrete action space is used to handle both discrete and continuous actions, with continuous actions like acceleration or deceleration being discretized into a range of discrete values. However, this approach of approximating continuous actions with discrete actions has its limitations. It can result in the loss of the natural structure inherent in the continuous action space. Moreover, to achieve an approximation that is precise enough, a considerably large set of discrete actions is often required, which can complicate the model and reduce its efficiency \citep{khamassi2017active,hausknecht2015deep,masson2016reinforcement}.

Therefore, introducing an advanced RL-based algorithm with a hybrid action space that can integrate lane-changing and car-following control is essential for the safe and efficient automated lane change. This paper adopted a novel Deep RL framework, parametrized deep Q-network (P-DQN), to integrate the lane-changing and car-following behavior. The P-DQN can directly work on the discrete-continuous hybrid action space without discretizing continuous action and it has the merits of both deep deterministic policy gradient (DDPG) and deep Q-network (DQN) [33]. DDPG can output continuous actions by learning a deterministic policy network, while DQN is a popular choice for discrete decision-making problem \citep{lin2020comparison,huang2021usv}. 

The main contributions of this paper are i) integrated control of lane changing and car-following behavior to maintain a balance between traffic safety, comfort, and efficiency; ii) implementation of P-DQN with a hybrid action space, facilitating the completion of lane-changing involving four sub-objectives (e.g., avoiding collisions, car-following, lane-keeping, lane-changing). The rest of this paper is organized as follows. Section 2 described the controller model for the integrated control of the car-following and lane changing, including state space design, action space design, and reward functions design. Section 3 developed the P-DQN algorithm. Section 4 described the details of numerical simulated experiment and  the results of the P-DQN and MOBIL. Section 5 concludes this study.

\section{Model Development}
In this section, we present the development of a controller model for integrated car-following and lane-changing control. The automated vehicle is controlled using the RL algorithm P-DQN, which is trained based on the interactions between the environment and the RL agent. The environment provides the system state and updates it according to the agent's action. The RL agent receives the system state and outputs actions through a policy network, which is improved iteratively by the P-DQN algorithm based on reward functions. The system state and its evolution are detailed in Section 2.1, while the action space and reward function design for the P-DQN algorithm are discussed in Sections 2.2 and 2.3, respectively. Section 2.4 provides a more in-depth explanation of the P-DQN algorithm based on the previously defined state space, action space, and reward functions.

\subsection{State Space}
\label{subsec1}
\begin{figure}[t]
    \centering
    \includegraphics[width=0.7\linewidth]{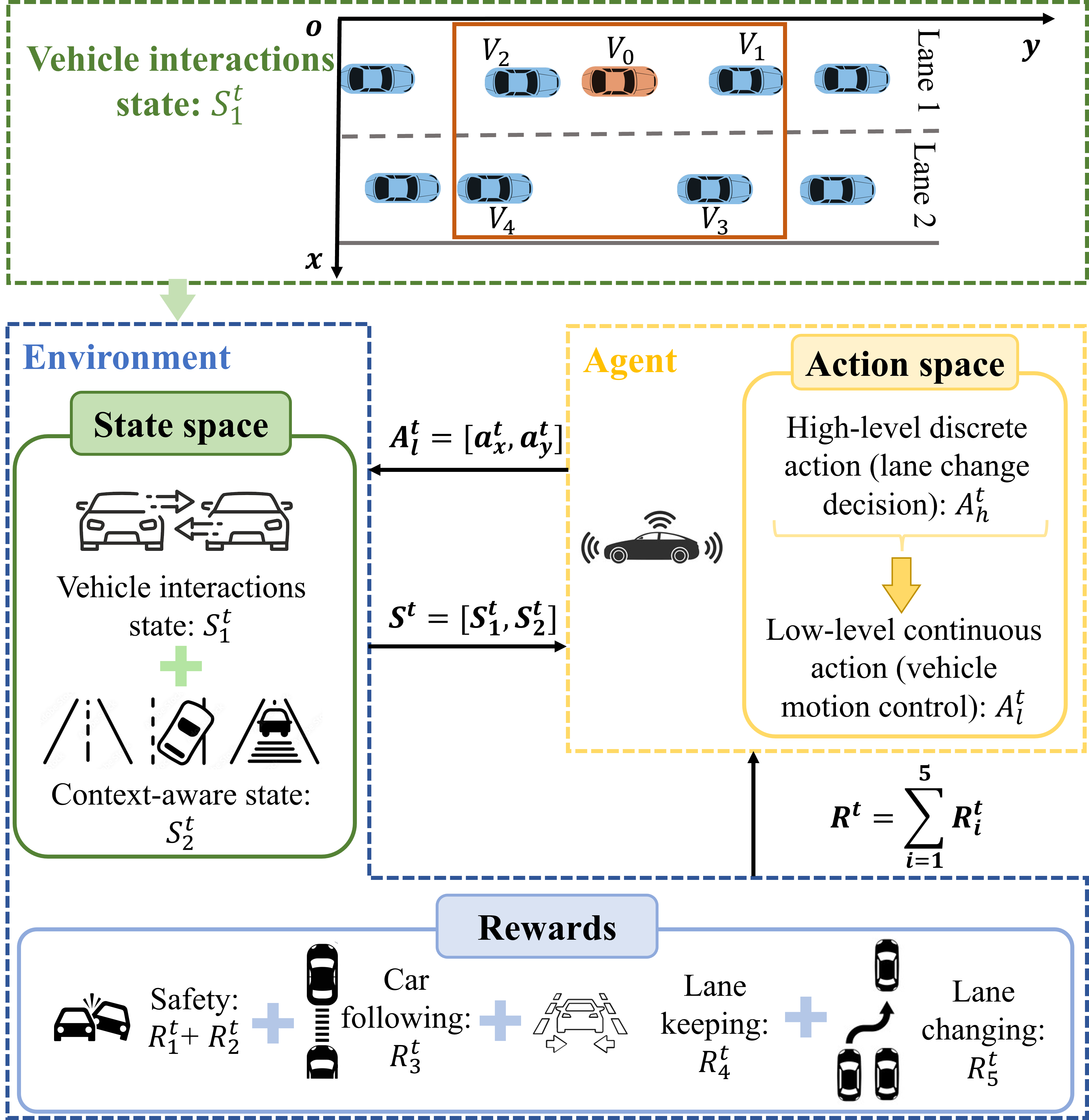}
    \caption{RL model development}
    \label{fig:2}
\end{figure}

The system state for car-following and lane-changing control typically includes vehicle position, velocity, spacing deviation, and velocity difference. Car-following control requires the state of the leading vehicle, while lane-changing control requires the state of surrounding vehicles in adjacent lanes as well as lane positions. Since this study integrates both car-following and lane-changing control, the state space consists of two main components: the state of the vehicles and the state of the road (e.g., road boundary and lane centerline), which is shown in Fig. \ref{fig:2}. The combined state space is defined as:

\begin{equation}
S^t = (S_1^t, S_2^t)
\end{equation}

where $S_1^t$ represents the vehicle interactions state, defined as $S_1^t = [s_1^t, s_2^t, s_3^t, s_4^t]$, where $s_i^t$ denotes the state of surrounding vehicle $i$ with $i \in \{1, 2, 3, 4\}$. As shown in Fig. \ref{fig:2}, the ego vehicle collects the states of four surrounding vehicles: $V_1$ (leading vehicle), $V_2$ (following vehicle), $V_3$ (leading vehicle in the adjacent lane), and $V_4$ (following vehicle in the adjacent lane). The state $s_i^t$ for each surrounding vehicle is defined as:

\begin{equation}
s_i^t = [\Delta x_i^t, \Delta y_i^t, \Delta v_{x,i}^t, \Delta v_{y,i}^t, \Delta n_i^t]^T
\end{equation}

where $\Delta x_i^t = x_0^t - x_i^t$ represents the lateral position difference between the ego vehicle ($i=0$) and vehicle $i$; $\Delta y_i^t = y_0^t - y_i^t$ denotes the longitudinal position difference; $\Delta v_{x,i}^t = v_{x,0}^t - v_{x,i}^t$  denotes the lateral velocity difference; $\Delta v_{y,i}^t = v_{y,0}^t-v_{y,i}^t$ denotes the longitudinal velocity difference; $\Delta n_i^t = n_0^t - n_i^t$ denotes the lane index difference. The state variables are updated based on the following dynamics:

\begin{equation}
v_{x,i}^{t+1} = v_{x,i}^t + a_{x,i}^t \Delta t
\end{equation}
\begin{equation}
x_i^{t+1} = x_i^t + v_{x,i}^t \Delta t + \frac{1}{2} a_{x,i}^t (\Delta t)^2
\end{equation}
\begin{equation}
n_i^{t} = \left\lfloor \frac{x_i^{t}}{L} \right\rfloor + 1
\end{equation}

where $\Delta t$ denotes the time step and $L$ is the lane width. The lateral and longitudinal velocities are constrained by predefined bounds. Positive values for lateral velocity indicate a right turn, and negative values indicate a left turn. The context-aware state, denoted by $S_2^t$, includes the ego vehicle's position relative to lane boundaries and the ideal lane centerline:

\begin{equation}
S_2^t = [\Delta p_r^t, \Delta p_l^t, \Delta p_c^t, \Delta n_0^t]^T
\end{equation}

where $\Delta p_r^t = x_0^t$ represents the lateral distance to the upper road boundary; $\Delta p_l^t = 2L - x_0^t$ denotes the lateral distance to the lower road boundary; $\Delta p_c^t = |x_0^t - p_c^{*,t}|$ represents the distance to the ideal lane centerline; $\Delta n_0^t = n_0^t - n_0^{*,t}$ denotes the lane index difference between the current lane and the ideal lane. The ideal lane ($n_0^{*,t}$) is chosen to optimize the ego vehicle’s speed. For instance, if the adjacent lane’s leading vehicle ($V_3$) is faster than the current lane’s leading vehicle ($V_1$), the ego vehicle shifts to lane 2, with $V_3$ becoming the new leading vehicle. The ideal lane index and centerline can be computed as:

\begin{equation}
n_0^{*,t} = \begin{cases}
n_0^t, & \text{if } v_{y,3}^t \leq v_{y,1}^t \\
n_3^t, & \text{otherwise}
\end{cases}
\end{equation}
\begin{equation}
p_c^{*,t} = (n_0^{*,t} - 1) \cdot L + \frac{L}{2}
\end{equation}

where $v_{y,3}^t$ and $v_{y,1}^t$ represent the longitudinal velocities of $V_3$ and $V_1$, respectively. If no vehicle is present in the adjacent lane, $v_{y,3}^t$ is set to the free-flow speed.

\subsection{Action Space}
\label{subsec2}
In this study, the P-DQN algorithm operates using a hybrid action space that includes one discrete action, denoted by $k^t$, which represents the lane-change decision at time step $t$, and two continuous actions, $a_{x,0}^t$ and $a_{y,0}^t$. The continuous actions, $a_{x,0}^t$ and $a_{y,0}^t$, control the lateral and longitudinal accelerations of the ego vehicle at time step $t$, respectively. The action space is defined as:
\begin{equation}
A^t = [A^t_h, A^t_l]^T
\end{equation}
where $A^t_h = k^t$ represents the high-level discrete action, and $A^t_l = [a_{x,0}^t, a_{y,0}^t]^T$ represents the low-level continuous actions. While both high-level and low-level actions are defined, only the low-level continuous actions are directly applied to the environment. The high-level discrete action guides the decision-making process by influencing the selection of low-level continuous action values. The discrete action $k^t$ has three possible options:
\begin{equation}
k^t \in \{0, 1, 2\}
\end{equation}
where $k^t = 0$ represents changing to the left lane; $k^t = 1$ represents staying in the current lane; $k^t = 2$ represents changing to the right lane. Each discrete action defines specific ranges for the continuous actions. All discrete actions share the same range for longitudinal acceleration, while the lateral acceleration range varies depending on the discrete action chosen. Table \ref{tab:1} illustrates the acceleration boundaries, where $a_{x}^{b}$ and $a_{y}^{b}$ denote the boundary values for lateral and longitudinal accelerations, respectively. For $k^t = 0$, corresponding to a left lane change, the lateral acceleration $a_{x,0}^t$ is constrained to negative values, ensuring the vehicle moves toward the left. For $k^t = 2$, corresponding to a right lane change, the lateral acceleration is positive, indicating movement toward the right. For $k^t = 1$, which signifies staying in the current lane, the lateral acceleration $a_{x,0}^t$ should ideally be zero. However, since the ego vehicle may not be perfectly centered in the lane, a small tolerance $\epsilon$ is allowed, enabling minor lateral adjustments to maintain lane position. Thus, the continuous actions allow fine control over both the longitudinal and lateral accelerations, while the discrete actions govern high-level lane-changing decisions, effectively integrating both components in the hybrid action space framework.

\begin{table}[htbp]
    \centering
    \caption{Hybrid action space design} 
    \label{tab:1} 
    \begin{tabular}{l l l l}
        \hline
        Action & $k^t$ & $a_{y,0}^t$ ($m/s^{2}$)& $a_{x,0}^t$ ($m/s^{2}$)\\
        \hline
        & 0 &  & $[-a_{x}^{b}, -\epsilon]$\\
        Value & 1 & $[-a_{y}^{b}, a_{y}^{b}]$& $[-\epsilon, \epsilon]$ \\
        & 2 &  & $[\epsilon, a_{x}^{b}]$\\
        \hline
    \end{tabular}
\end{table}

\subsection{Reward functions}
\label{subsec3}
The reward functions are designed to align with the learning objectives of the integrated control problem, which encompasses both lane-changing and car-following behaviors. Since these behaviors involve multiple objectives, the reward functions are structured to address key aspects such as safety, car-following performance, lane-keeping, and lane-changing efficiency.

\subsection*{a) Safety Reward}
The safety reward consists of two parts: avoiding collisions with surrounding vehicles and avoiding collisions with the road boundary. Safety reward stems from the proximity to surrounding vehicles and the road boundary. As shown in Eq. \eqref{eq:collision_risk}, to avoid a collision with surrounding vehicles, a rational penalty is given when the lateral distance $\Delta x_i^t$ and longitudinal distance $\Delta y_i^t$ between the ego vehicle and vehicle $i$ are simultaneously small. Distance thresholds $\delta = [\delta_1, \delta_2, \delta_3]$ are defined to evaluate if there is a risk of collision. $R_{1,i}^t$ represents the reward for the vehicle collisions between the ego vehicle and surrounding vehicle $i$, expressed as follows:
\begin{equation}
\begin{split}
R_{1,i}^t = 
\begin{cases} 
\max \left( \log_2 \left( \frac{\Delta y_i^t}{\delta_2}\right), R^{b} \right) + \\
\max \left( \log_2 \left( \frac{\Delta x_i^t}{\delta_1} \right), R^{b} \right), & \text{if } \Delta y_i^t < \delta_2 \text{ and } \Delta x_i^t < \delta_1 \\
0, & \text{else}
\end{cases}
\end{split}
\label{eq:collision_risk}
\end{equation}

\begin{equation}
R_{1}^t = \sum_{i=1}^{4} R_{1,i}^t
\label{eq:sum_safety_reward}
\end{equation}

where $R^{b}$ is the boundary value of $R_{1,i}^t$. Previous research demonstrated infinitesimal reward values make the RL algorithm difficult to converge \citep{zhou2021multi}. Thus, this study established bounded values for reward functions. $R_{1}^t$ denotes the sum of the collision reward of four surrounding vehicles. $\delta_1$ and $\delta_2$ represent the lateral and longitudinal thresholds of the collision risk. Moreover, the ego vehicle must avoid collisions with the road boundary, which can be written as:

\begin{equation}
R_{2}^t = \begin{cases} 
\max \left( \log_2 \left( \frac{\Delta p_l^t}{\delta_3} \right), R^{b} \right), & \text{if } \Delta p_l^t < \delta_3 \\
\max \left( \log_2 \left( \frac{\Delta p_u^t}{\delta_3} \right), R^{b} \right), & \text{if } \Delta p_u^t < \delta_3 \\
0, & \text{else}
\end{cases}
\end{equation}

where $\Delta p_l^t$ and $\Delta p_u^t$ represent the ego vehicle's distance to the lower and upper boundary of the road. When the distance between the ego vehicle and the boundary of the road is less than $\delta_3$, there is a collision risk and $R_{2}^t$ will be negative. For the safety rewards, when the distance is smaller, the values of the reward will be smaller. The values of the safety reward will always be negative. 

\subsection*{b) Car-Following Reward}
The objective of the car-following reward $R_{3}^t$ is to enable the ego vehicle maintains a constant time gap with its leading vehicle $V_1$. The reward function is written as:

\begin{equation}
R_{3}^t = \max \left( R^{b}, - \left( v_{y,0}^{*,t} - v_{y,0}^t \right)^2 \right)
\label{eq:car_following_reward}
\end{equation}

\begin{equation}
v^{*,t}_{y,0} = 
\begin{cases}
v_y^{f}, & \text{if } \Delta y_1^t \geq d \\
\max \left( 0, \left( \Delta y_1^t - d_s \right) / \tau \right), & \text{else}
\end{cases}
\label{eq:desired_velocity}
\end{equation}

\begin{equation}
d = \tau \cdot v_y^f + d_s
\label{eq:headway_distance}
\end{equation}

where $R^{b}$ is the boundary value of $R_{3}^t$, $v_{y,0}^t$ is the longitudinal velocity of the ego vehicle, $v^{*,t}_{y,0}$ is the desired velocity of the ego vehicle which is calculated by Eq. \eqref{eq:desired_velocity}. $\Delta y_1^t$ is the longitudinal distance between the ego vehicle and the leading vehicle $V_1$. $d$ represents the desired spacing; when $\Delta y_1^t$ is larger than $d$, the ego vehicle can travel at the free-flow speed $v_y^f$. $\tau$ is the desired time gap between two successive vehicles, and $d_s$ is the standstill distance. When $\Delta y_1^t$ is less than the desired spacing, the ego vehicle's velocity is smaller than  $v_y^{f}$ which makes the ego vehicle has a desired time gap with the leading vehicle.

\subsection*{c) Keeping-Lane Reward}
The keeping-lane reward $R_{4}^t$ stems from the ego vehicle's deviation from the lane centerline. This reward encourages the ego vehicle to drive along the centerline of the road while it is not changing lanes. $R_{4}^t$ is expressed as follows:

\begin{equation}
R_{4}^t = - \left( x_0^t - p_c^{*,t} \right)^2
\label{eq:keeping_lane_reward}
\end{equation}

where $x_0^t$ is the lateral position of the ego vehicle, $ p_c^{*,t}$ is the position of the ideal lane's centerline. $R_{4}^t$ is a negative reward, decreasing as the difference between the ego vehicle's lateral position and the lane centerline increases.

\subsection*{d) Lane-Changing Reward}
The reward for lane-changing is quantified by evaluating the vehicle's alignment with the ideal lane. This assessment is conducted through the analysis of the vehicle's lateral velocity direction in relation to the direction of the ideal lane, as described by the following equation:

\begin{equation}
R_{5}^t =
\begin{cases}
R^{b}, & \text{if } v_{x,0}^t \cdot (n_0^{*,t} - n_{0}^t) > 0 \\
-R^{b}, & \text{if } v_{x,0}^t \cdot (n_0^{*,t} - n_{0}^t) < 0
\end{cases}
\label{eq:lane_changing_reward}
\end{equation}

where $v_{x,0}^t$ denotes the ego vehicle's lateral velocity, $n_{0}^t$ denotes the index of the lane where the ego vehicle is located, and $n_0^{*,t}$ is the index of the ideal lane. Different signs of $v_{x,0}^t$ denote the ego vehicle is driving towards distinct directions as explained in previous section.   $R^{b}$ represents the boundary value for the lane-changing reward. The lane-changing reward indicates that when the ego vehicle is moving toward the ideal lane, there is a positive reward, and conversely, a negative reward. As shown in the following equation, the proposed algorithm improves its performance by maximizing the cumulative sum of the rewards obtained. The details of parameters in all reward functions are shown in Table \ref{tab:2}.

\begin{equation}
R_{total}^t = R_{1}^t + R_{2}^t + R_{3}^t + R_{4}^t + R_{5}^t
\label{eq:total_reward}
\end{equation}

\begin{table}[htbp]
\centering
\caption{Parameter values for reward functions}
\label{tab:2}
\begin{tabular}{ll|ll}
\hline
Parameters & Value & Parameters & Value \\
\hline
$\delta_1$ & 2.8 & $R^{b}$ & -20 \\
$\delta_2$ & 14 & $a_{x}^b$& 0.3 \\
$\delta_3$ & 1.5 & $a_{y}^b$& 3 \\
$\tau$ & 2 & $d_s$ & 10\\
$\epsilon$& 0.01& $L$ & 3.75 \\
\hline
\end{tabular}
\end{table}

\section{P-DQN algorithm}
In this section, the P-DQN algorithm is described. P-DQN is developed based on Q-learning. Both algorithms are value-based algorithm, focusing on learning the action-value function network. Q-learning updates the action-value function as follows:
\begin{equation}
Q(x^t, a^t) \leftarrow Q(x^t, a^t) + \alpha \left[ r^{t+1} + \gamma \max_{a'} Q(x^{t+1}, a') - Q(x^t, a^t) \right]
\label{eq:q_learning_update}
\end{equation}
where $Q(x^t, a^t)$ is the action-value function at time step $t$, $r^{t+1}$ denotes the reward received after taking action $a'$ at state $s^{t+1}$. $\alpha$ is the soft learning rate, $\gamma$ is the discount factor. The learning objective of the Q-learning can be expressed as follows:
\begin{equation}
\pi_{t+1} = \arg\max_{a'} Q(x^{t+1}, a')
\label{eq:11}
\end{equation}
where $\pi_{t+1}$ is the policy in Q-learning, $Q(x^t, a^t)$ is expressed as a table. When the state space is continuous, DQN approximated $Q(x^t, a^t)$ by a neural network: $Q(x^t, a^t; \theta)$, where $\theta$ is the network weights. The DQN's action-value function updating process can be expressed as the Bellman expectation equation:

\begin{equation}
Q(x^t, a^t) = \mathbb{E} \left[ r^{t+1} + \gamma \max_{a'} Q(x^{t+1}, a'; \theta) \mid x^t, a^t \right]
\label{eq:12}
\end{equation}
where $\mathbb{E}$ denotes the expected value, $r^{t+1}$ denotes the immediate reward received after taking action $a'$ in state $x^{t+1}$, $\gamma \max_{a'} Q(x^{t+1}, a')$ denotes the discounted best further reward. DQN updates the neural network weights by the least square loss function:

\begin{equation}
L(\theta) = \mathbb{E} \left[ (r^{t+1} + \gamma \max_{a'} Q(x^{t+1}, a'; \theta) - Q(x^t, a^t; \theta))^2 \right]
\end{equation}

In this study, the action space is a hybrid discrete-continuous action space, the action-value function $Q(x^t, a^t)$ can be expressed as a parametrized form: $Q(x^t, a^t) = Q(S^t, k^t, A^t_l)$, where $S^t = (S^t_1,S^t_2)$ is the system state,  $k^t \in \{0, 1, 2\}$ denotes the high-level discrete action, $A^t_l=[a_{1,0}^t,a_{2,0}^t]^T$ represents the low-level continuous action. The $A^t_l$ is the parameter of the $k^t$. Then the Bellman expectation equation revised as:

\begin{equation}
Q(S^t, k^t, A^t_l) = \mathbb{E}[R_{total}^{t+1 }+ \gamma \max_{k'}\sup_{A'_l} Q(S^{t+1}, k', A'_l) \mid S^t, k^t, A^t_l]
\label{eq:pdqn bellman}
\end{equation}

where $R_{total}^{t+1 }$ denotes the immediate reward received after taking action $k', A'_l$ in state $S^{t+1}$. $R_{total}^{t+1 }$ is calculated by Eq. \eqref{eq:total_reward}. To solve this Bellman equation, P-DQN utilizes a deterministic policy network $A_k^Q (S^{t+1}, k^{t+1};\omega)$ to approximate the function $\sup_{A'_l} Q(S^{t+1}, k^{t+1}, A'_l)$, where $\omega$ represents the weights of the network. Then the Eq. \eqref{eq:pdqn bellman} is revised as:
\begin{equation}
Q(S^t, k^t, A^t_l) = \mathbb{E}[R_{total}^{t+1 }+ \gamma \max_{k'} Q(S^{t+1}, k', A_k^Q (S^{t+1}, k^{t+1};\omega)) \mid S^t, k^t, A^t_l]
\label{eq:newpdqn bellman}
\end{equation}
which is similar to the Bellman equation of the DQN, as shown in Eq. \eqref{eq:12}. Then, similar to the DQN, the  Eq. \eqref{eq:newpdqn bellman} is approximated by a value neural network $Q(S^t, k^t, A_k^Q (S^{t}, k^{t};\omega);\theta)$, where $\theta$ is the weights of the value network. Then the network weights $\theta$ and $\omega$ are estimated by minimizing the least-squares loss functions via gradient descent. Specially, the least-squares loss function for $\theta$  and $\omega$ can be written as:

\begin{equation}
L(\theta) = \left[ Q(S^t, k^t,A^t_l; \theta) - y^t \right]^2
\end{equation}
\begin{equation}
L(\omega) = - \sum_{k=0}^{2}  Q(S^t, k^t, A_k^Q (S^{t}, k^{t};\omega);\theta) 
\end{equation}
where $y^t$ denotes the expected further reward, it can be expressed as:
\begin{equation}
y^t =R_{total}^{t+1 }+ \gamma \max_{k'} Q(S^{t+1}, k', A_k^Q (S^{t+1}, k^{t+1};\omega); \theta)
\label{eq:further reward}
\end{equation}

\begin{figure}[h] 
    \centering
    \begin{minipage}{\textwidth}
        \begin{algorithm}[H] 
            \caption{Parametrized Deep Q-Network (P-DQN)}
            \begin{algorithmic}[1]
                \State Input: network weights learning rate: $\alpha$, $\beta$, exploration rate: $\epsilon$, minibatch size: $B$, distribution: $\varepsilon$, termination: $\vartheta$
                \State Initialize network weights: $\omega$, $\theta$
                \For{each step t}
                    \State Compute continuous action for each $k$: $ A_l \leftarrow A_k^Q (S^{t+1}, k^{t+1};\omega)$
                    \State Select discrete action and corresponding continuous action. $k'$ is selected according to the $\epsilon$-greedy policy. $k' = \max_{k'}Q(S^{t+1}, k', A'_l)$
                    \State $k'= 
                        \begin{cases}
                        \text{a sample from distribution } \varepsilon, & \text{with probability } \epsilon \\
                        \arg\max_{k'}Q(S^{t+1}, k', A'_l), & \text{with probability } 1-\epsilon
                        \end{cases}$
                    \State Execute action $A^t =[k',A'_l ] $, get a reward $R^t_{total}$, go to the next state: $S^{t+1}$
                    \State Store experience $(S^t, A^t, R^t_{total}, S^{t+1})$ into replay buffer
                    \State Randomly sample $B$ rows: $ \{S^t, A^t, R^t_{total}, S^{t+1}\}$ from replay buffer
                    \State Compute learning target $y^t$
                    \State $y^t = 
                        \begin{cases}
                        R^{t+1}_{total}, & \text{if } \vartheta = 1 \\
                        R^{t+1}_{total} + \gamma \max_{k'} Q(S^{t+1}, k', A_k^Q (S^{t+1}, k^{t+1};\omega); \theta), & \text{else}
                        \end{cases}$
                    \State Utilize $B$ rows' data $\{y^t, S^t, A^t\}$ to compute the stochastic gradient $\nabla_{\omega} L(\omega)$ and $\nabla_{\theta} L(\theta)$
                    \State Update the weights: $\omega^{t+1} \leftarrow \omega^t - \alpha \nabla_{\omega} L(\omega)$ and $\theta^{t+1} \leftarrow \theta^t - \beta \nabla_{\theta} L(\theta)$
                    \If{$S^{t+1}$ is the terminal state or a collision occurs:}
                        \State update $\vartheta = 1$
                        \State \textbf{break}
                    \EndIf
                \EndFor
            \end{algorithmic}
        \end{algorithm}
    \end{minipage}
\end{figure}

The process of the P-DQN algorithm is demonstrated in the following table. The DQN outputs the discrete action based on the critic network. Different from the DQN, the P-DQN use one more actor network to output the continuous action. In terms of the network structure, the P-DQN is similar to the DDPG. Thus, the P-DQN requires training two networks. The networks are optimized by minimizing the loss functions using stochastic gradient methods. The learning rate is set as 0.0001. The target for the loss function is the expected reward in the further, as illustrated by Eq. \eqref{eq:further reward}. The discount factor for the further reward is set as 0.9. Moreover, there are five rewards in this study, the coefficients for the safety rewards are set to be larger than other rewards because the safety is the primary control objective. The coefficients for other rewards are tuned appropriately based on the performance of the P-DQN algorithm. In this study, the coefficient for the $R^t_1$ and $R^t_2$ is set as 20. The coefficients for $R^t_3$, $R^t_4$, and $R^t_5$ are set as:0.5, 3, 1. Because large reward values might cause an increase in computation time during reinforcement learning training, this study scales down the total reward value to one-thousandth of its original value during training \citep{ma2022lyapunov}.
The hyperparameters of the algorithm in this study are adjusted according to existing related studies \citep{shi2022deep,zhang2019simultaneous,xiong2018parametrized}.The batch size is set as 128. The exploration rate of the $\epsilon$-greedy policy is set as 0.05. The $\epsilon$-greedy policy can diverse the outputs of the RL algorithm and avert local optimal solution.

\begin{figure}[t]
    \centering
    \includegraphics[width=0.8\linewidth]{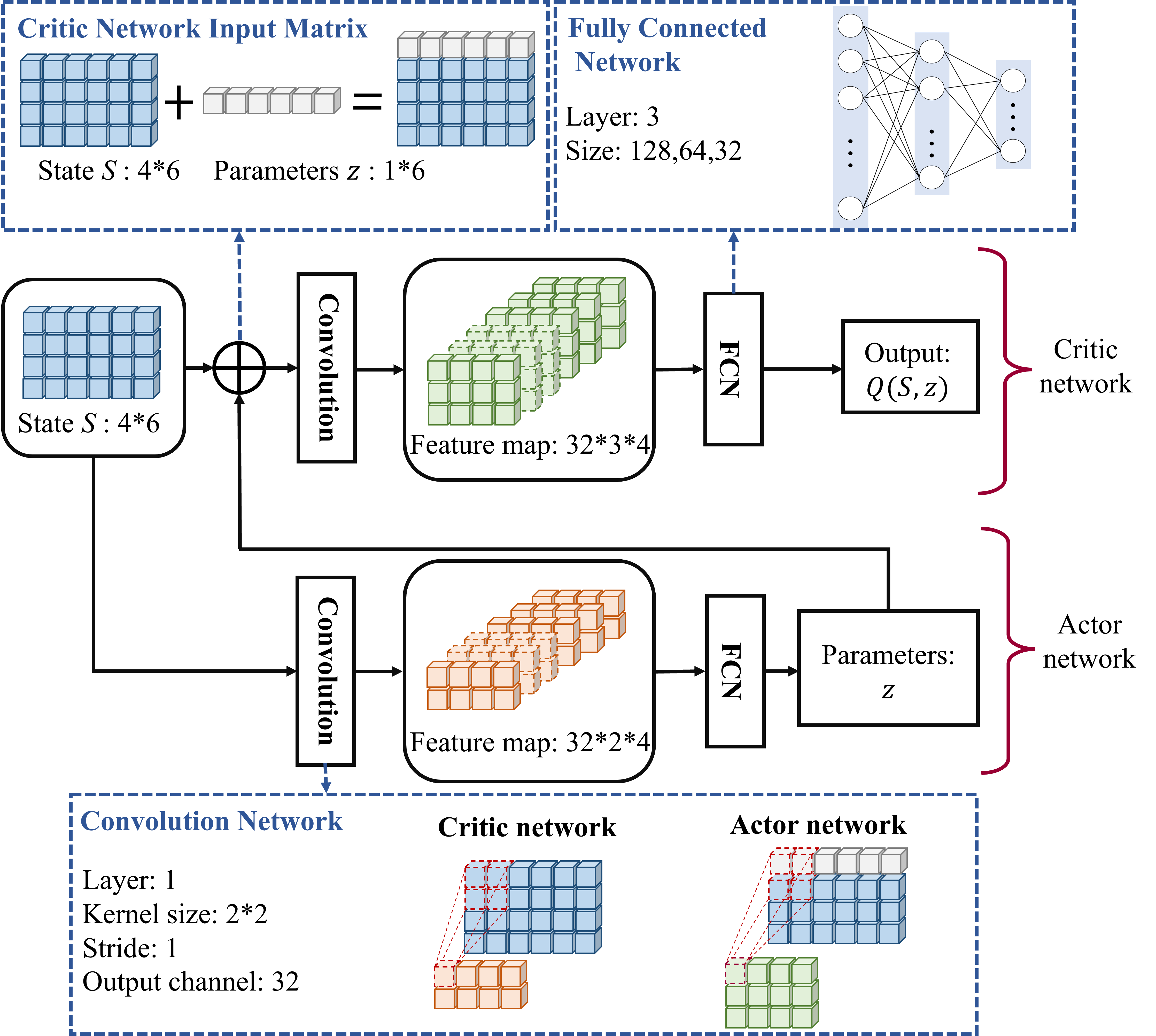}
    \caption{Neural network structure of the P-DQN}
    \label{fig:3}
\end{figure}

As discussed in Section 2, the state space of the integrated control of the car-following and lane changing consists of 24 variables.  $S^t_1$ represents the state of the four surrounding vehicles, which is defined as: $S^t_1 = [s^t_1, s^t_2, s^t_3, s^t_4]$. Each vehicle's state includes five variables.  $S^t_2$ represents the state of the road, which includes four variables. It is challenging to extract the most essential information among 24 variables. This study adds an additional convolutional layer to the original neural network to extract effective features from the input information. The neural network structure used is illustrated in Fig. \ref{fig:3}. The output of the critic network is utilized to select the discrete action, as shown in Eq. \eqref{eq:newpdqn bellman}. The output of the actor network is the parameters of the discrete action, which is also the low level continuous action. Finally, the termination conditions in simulation experiments is crucial, as improper termination conditions might lead to the premature end of these experiments. It may cause RL algorithm hard to converge, due to insufficient training. We set three termination conditions: (i) the ego vehicle collides with any of the surrounding vehicles; (ii) the ego vehicle crosses the road boundary line; (iii) any vehicle in the environment reaches the end of the simulation section.

\section{Numerical experiments and results}
\subsection{Numerical experiments}
\label{subsec1}
\begin{figure}[htbp]
    \centering
    \begin{subfigure}[b]{0.49\textwidth}
        \includegraphics[width=\textwidth]{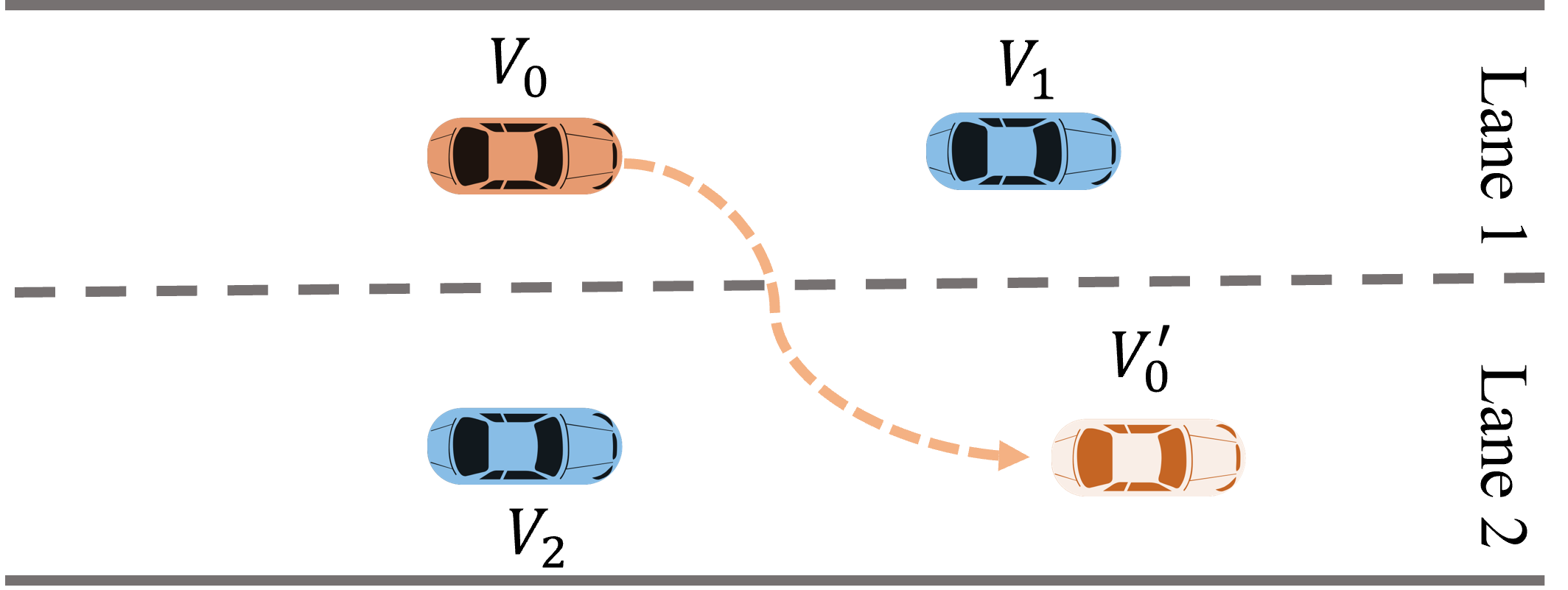}
        \caption{Scenario 1}
        \label{fig:4sub1}
    \end{subfigure}
    \hfill
    \begin{subfigure}[b]{0.49\textwidth}
        \includegraphics[width=\textwidth]{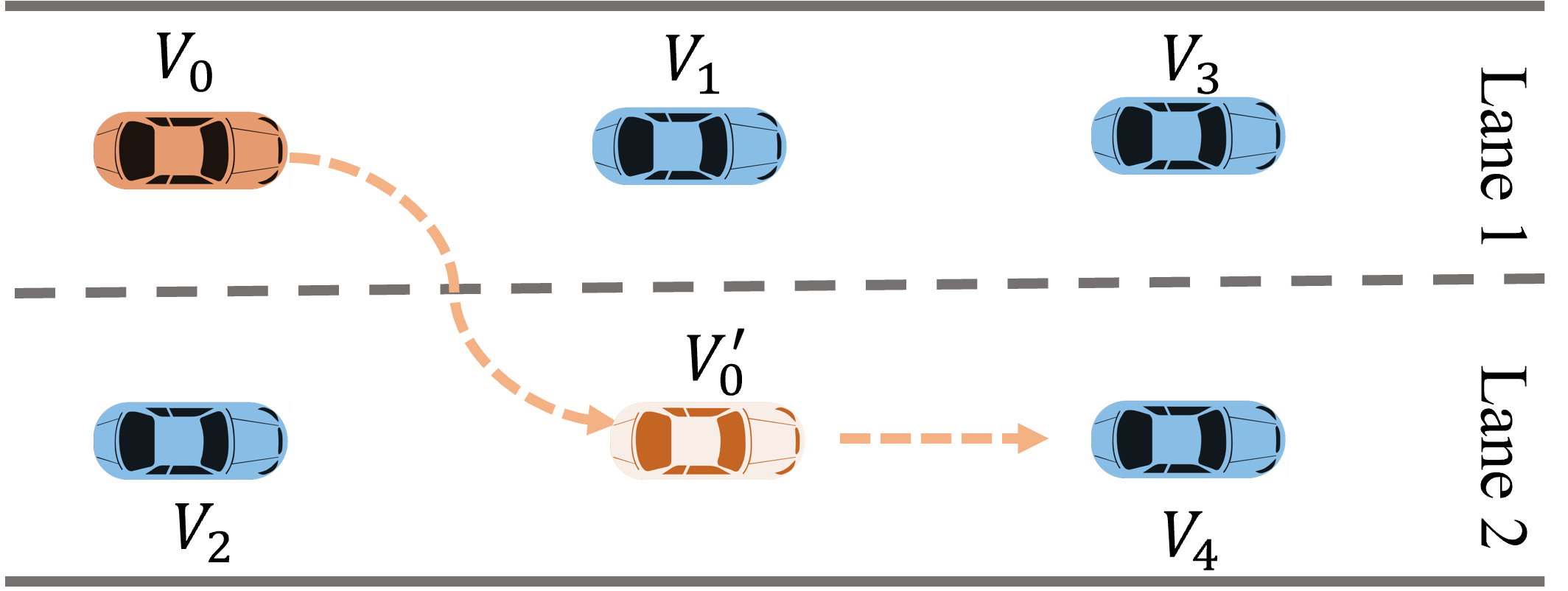}
        \caption{Scenario 2}
        \label{fig:4sub2}
    \end{subfigure}
    \hfill
    \begin{subfigure}[b]{0.49\textwidth}
        \includegraphics[width=\textwidth]{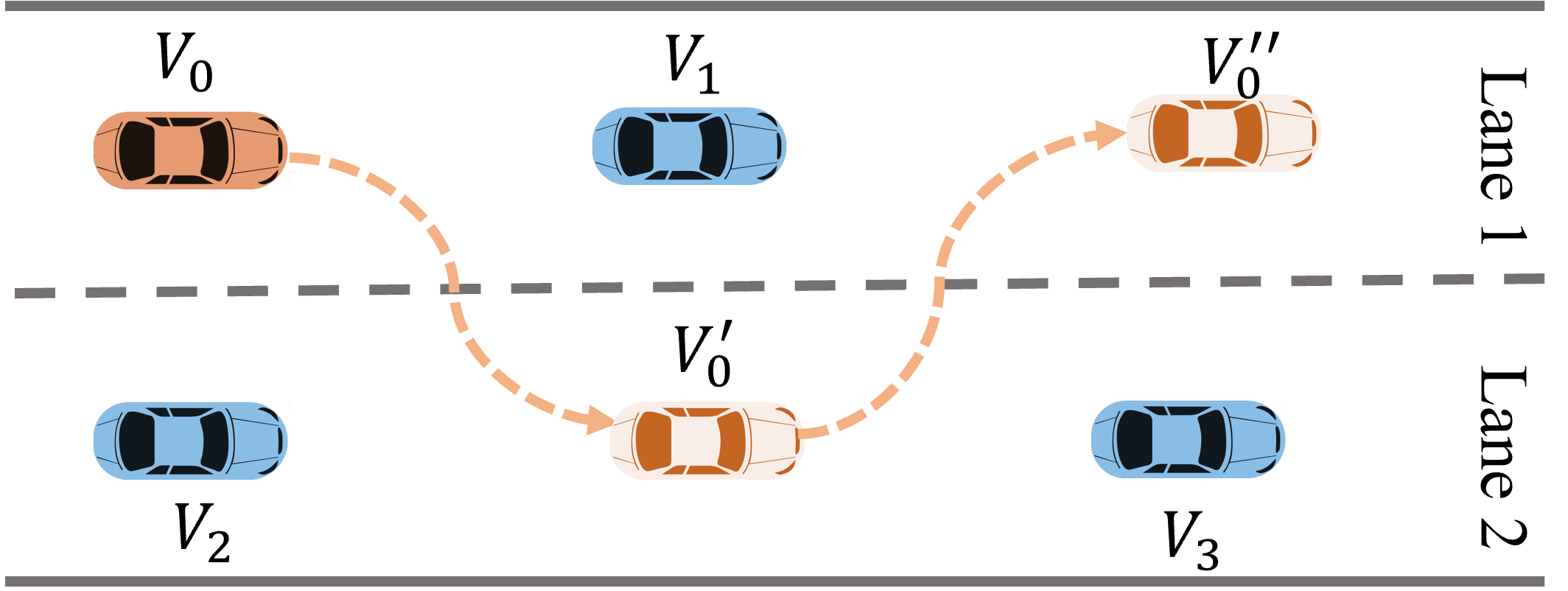}
        \caption{Scenario 3}
        \label{fig:4sub3}
    \end{subfigure}
    \hfill
    \begin{subfigure}[b]{0.49\textwidth}
        \includegraphics[width=\textwidth]{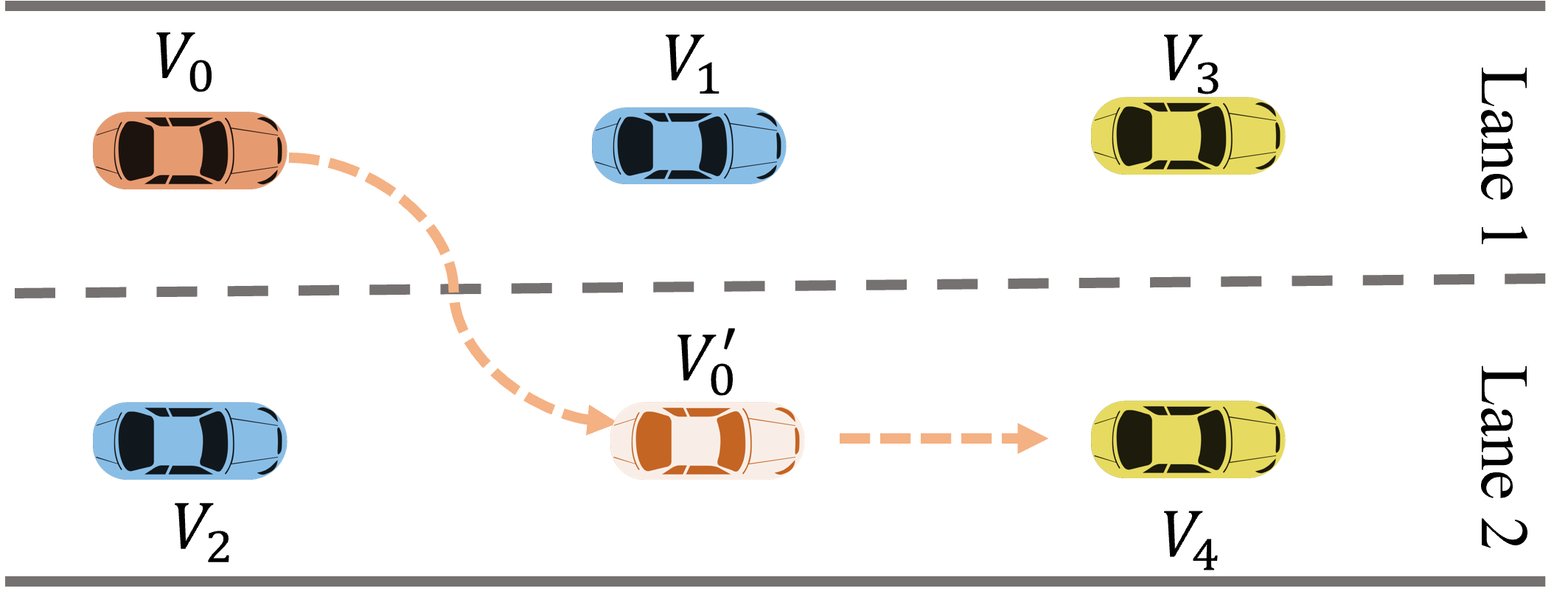}
        \caption{Scenario 4}
        \label{fig:4sub4}
    \end{subfigure}
    \hfill
    \caption{Scenarios setting of numerical experiments}
    \label{fig:4}
\end{figure}

As shown in Fig. \ref{fig:4}, this study designs four scenarios in numerical experiments to verify the performance of the P-DQN algorithm. The integrated control of the lane-changing and car-following behaviors includes four learning objectives: (i) The ego vehicle can avoid a collision with surrounding vehicles and the road boundary. (ii) The ego vehicle can achieve a higher driving efficiency through the automated lane change. (iii) The ego vehicle can complete the car-following and lane-keeping behaviors when it is not in a lane-changing process. Four scenarios vary in difficulty for achieving the stated objectives.
Table \ref{table:3} describes the vehicles’ initial state of four scenarios. As shown in Fig. \ref{fig:4}, in scenario 1, $V_1$ travels at a constant speed of 5 m/s. The ego vehicle has an initial speed of 5 m/s and starts at a location of 0 m in the longitudinal direction. $V_1$ starts at a location of 60 m. $V_1$ and the ego vehicle start on lane 1. There is a following vehicle $V_2$ behind the ego vehicle on lane 2. $V_2$ travels at a constant speed of 5 m/s and starts at a location of 0 m. Regarding learning objective (i), scenario 1’s ego vehicle should avoid a collision with $V_1$ and $V_2$. Regarding learning objective (ii), scenario 1’s ego vehicle should change from lane 1 to lane 2, then drive at free flow speed on lane 2. Regarding learning objective (iii), since there is not a leading vehicle on lane 2 in scenario 1, the ego vehicle will drive at free flow speed on lane 2. Different from scenario 1, scenario 2 has $V_3$ and $V_4$ in front of $V_1$. Thus, regarding learning objective (iii), after lane changing the ego vehicle in scenario 2 will try to keep a constant headway with $V_4$. Scenario 3 is more complicated, because the ego vehicle has to complete two lane change processes. In scenario 3, $V_3$ travels at a constant speed of 7 m/s and starts at a location of 120 m on lane 2. Regarding learning objective (ii), after changing from lane 1 to lane 2, the ego vehicle will try to change from lane 2 to lane 1 to overhead the $V_3$ and travel at free flow speed. As shown in Fig. \ref{fig:4sub4}, scenario 4 is designed to further demonstrate the P-DQN’s ability to learn car-following tasks. In contrast to scenario 2, leading vehicles $V_3$ and $V_4$ do not drive at a constant speed in scenario 4. Instead, they have a dynamic longitudinal acceleration. 

\begin{table}[htbp]
\centering
\caption{Scenario setting}
\label{table:3}
\begin{tabular}{llll}
\hline
Scenarios & Vehicles & Velocity (m/s) & Position (m) \\
\hline
1 & $V_0$ & 5 & $(L_w * 0.5, 0)$ \\
  & $V_1$ & 5 & $(L_w * 0.5, 50)$ \\
2 & $V_2$ & 5 & $(L_w * 1.5, 0)$ \\
  & $V_3$ & 6 & $(L_w * 0.5, 70)$ \\
  & $V_4$ & 6 & $(L_w * 1.5, 70)$ \\
3 & $V_3$ & 7 & $(L_w * 1.5, 120)$ \\
4 & $V_3$ & 10 & $(L_w * 0.5, 70)$ \\
  & $V_4$ & 10 & $(L_w * 1.5, 70)$ \\
\hline
\end{tabular}
\end{table}

\subsection{Results analysis}
\label{subsec2}
\subsubsection{P-DQN's results}
\begin{figure}[h]
    \centering
    \includegraphics[width=0.7\linewidth]{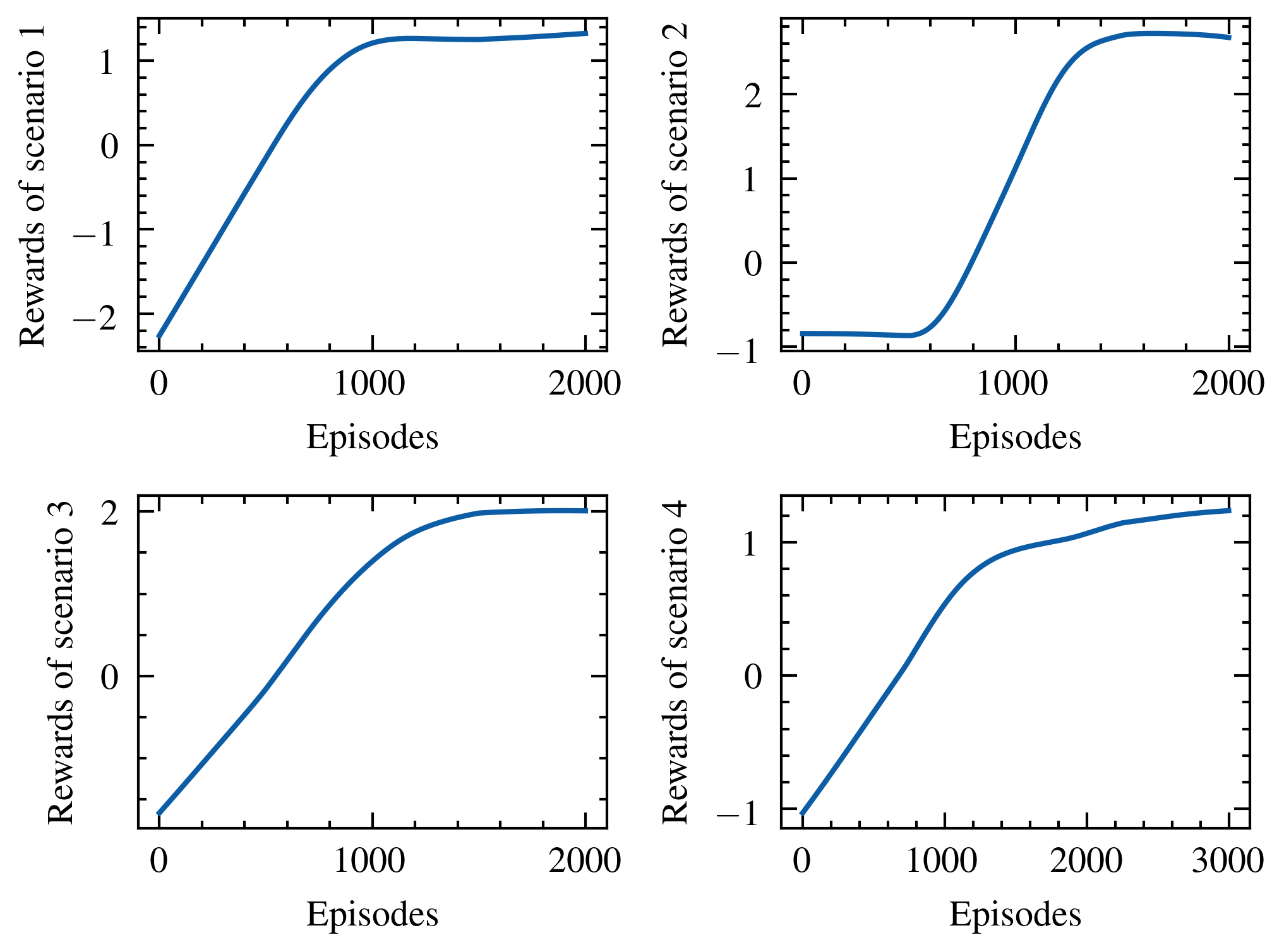}
    \caption{Illustration for total reward changes in four scenarios}
    \label{fig:5}
\end{figure}

The smoothed total reward of four scenarios is depicted in Fig. \ref{fig:5}. The P-DQN algorithm achieved convergence after 2,000 training episodes in scenarios 1 through 3. In contrast, scenario 4 necessitated 3,000 episodes for convergence, attributable to the variable speeds of the leading vehicle, which is more complex than other scenarios.

\begin{figure}[h]
    \centering
    \begin{minipage}{0.9\textwidth}
        \centering
        \begin{subfigure}[b]{0.48\textwidth}
            \includegraphics[width=\textwidth]{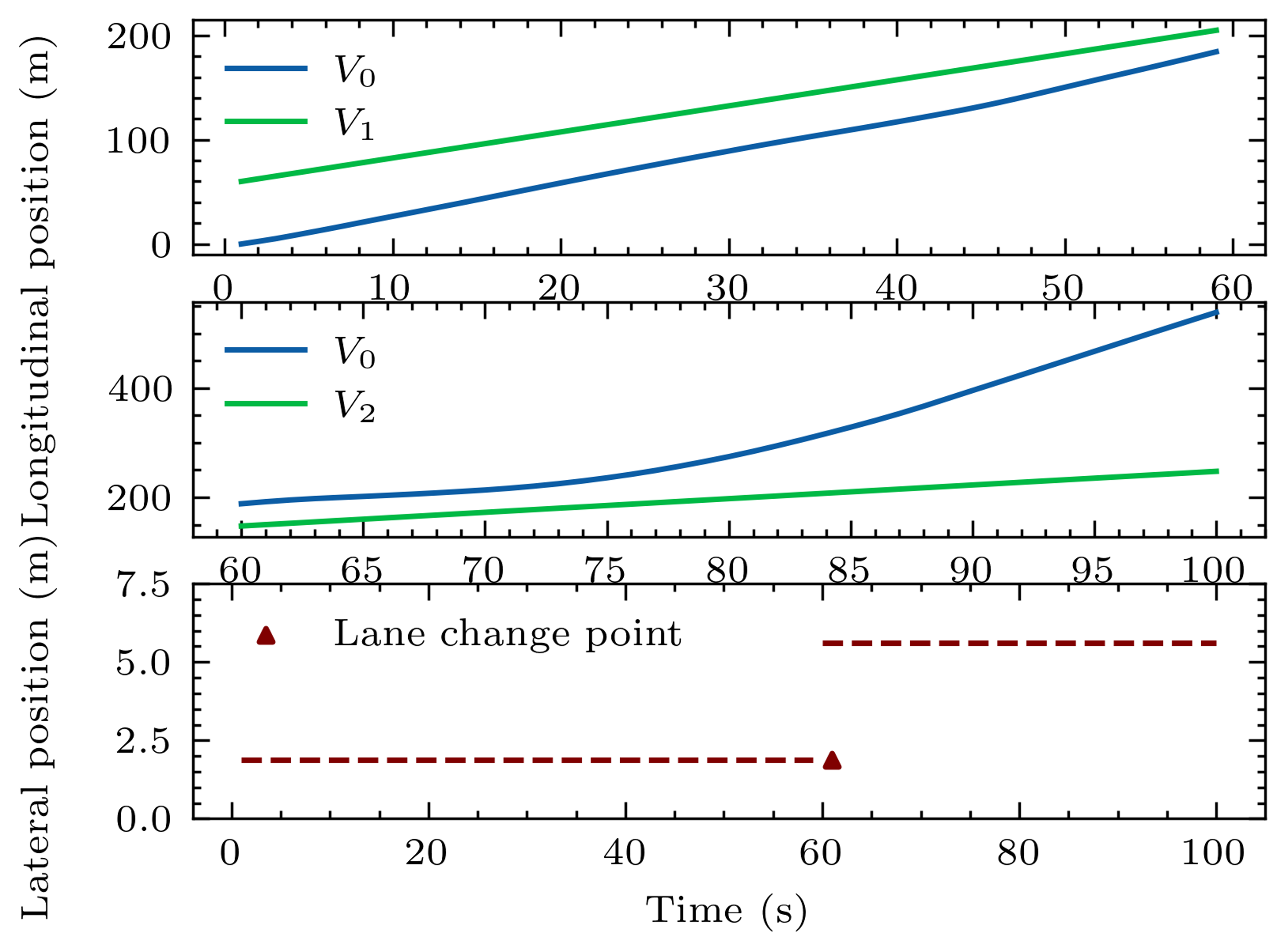}
            \caption{Scenario 1}
            \label{fig:6sub1}
        \end{subfigure}
        \hfill
        \begin{subfigure}[b]{0.48\textwidth}
            \includegraphics[width=\textwidth]{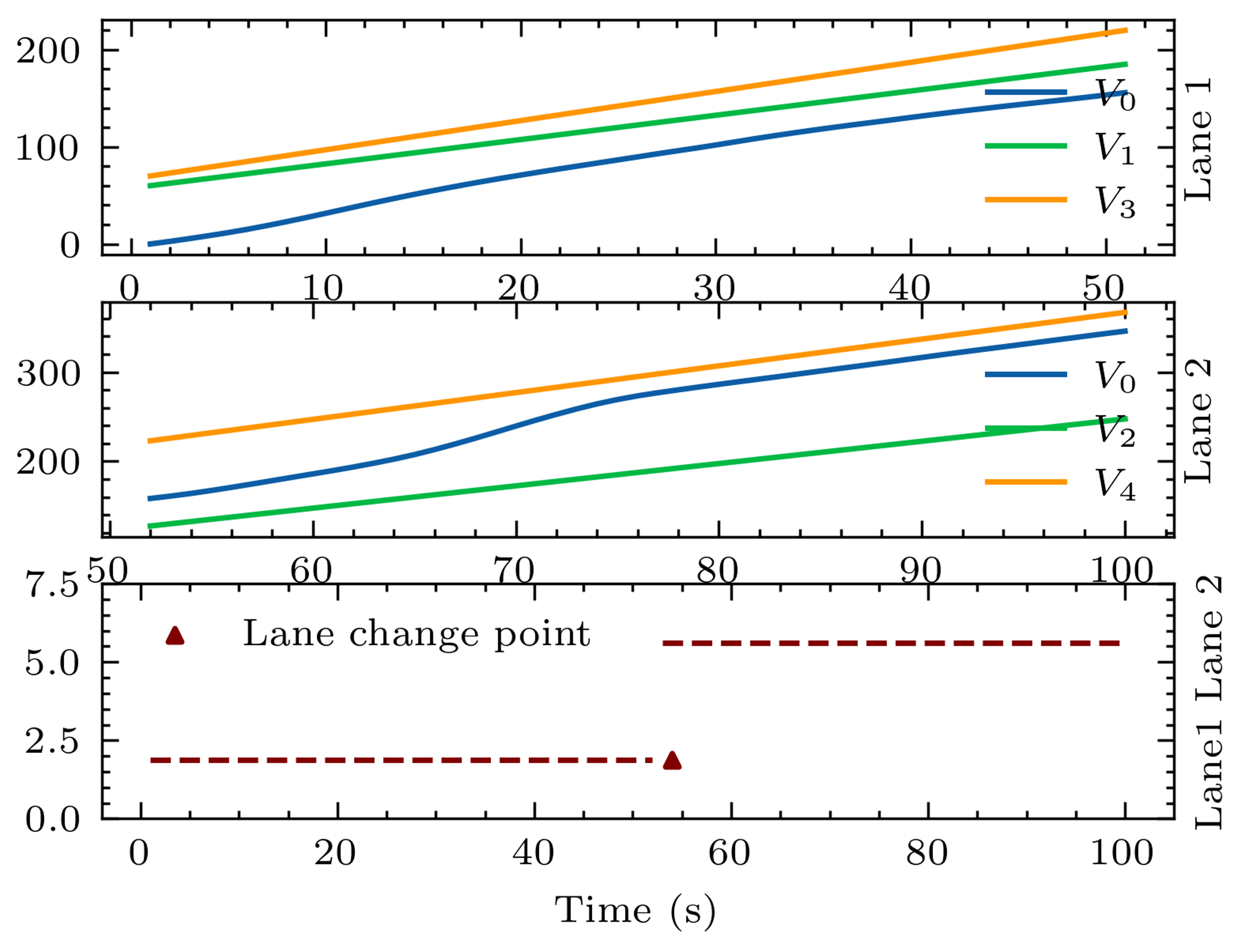}
            \caption{Scenario 2}
            \label{fig:6sub2}
        \end{subfigure}
    \end{minipage}
    
    \vspace{1em}
    
    \begin{minipage}{0.9\textwidth}
        \centering
        \begin{subfigure}[b]{0.48\textwidth}
            \includegraphics[width=\textwidth]{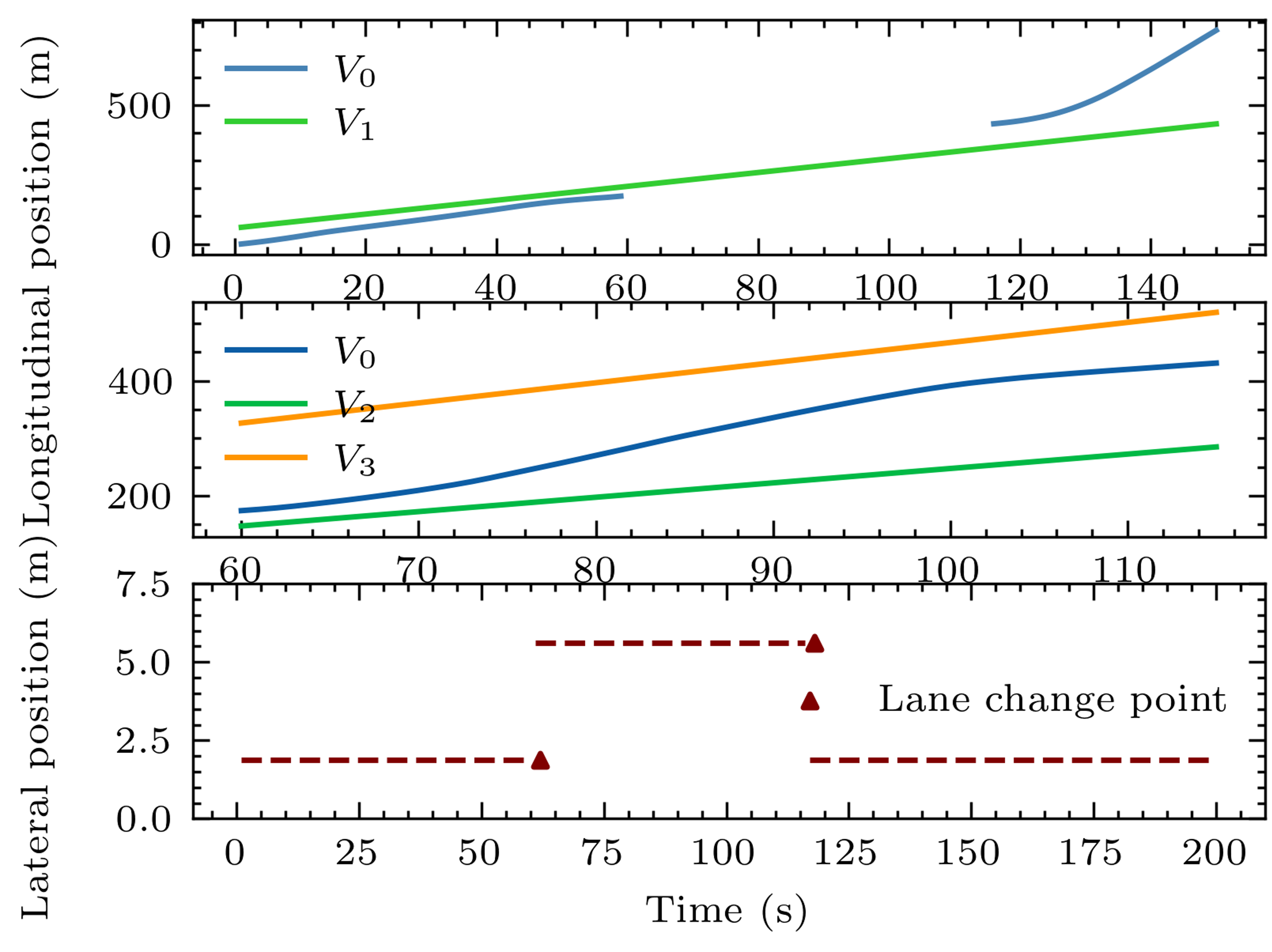}
            \caption{Scenario 3}
            \label{fig:6sub3}
        \end{subfigure}
        \hfill
        \begin{subfigure}[b]{0.48\textwidth}
            \includegraphics[width=\textwidth]{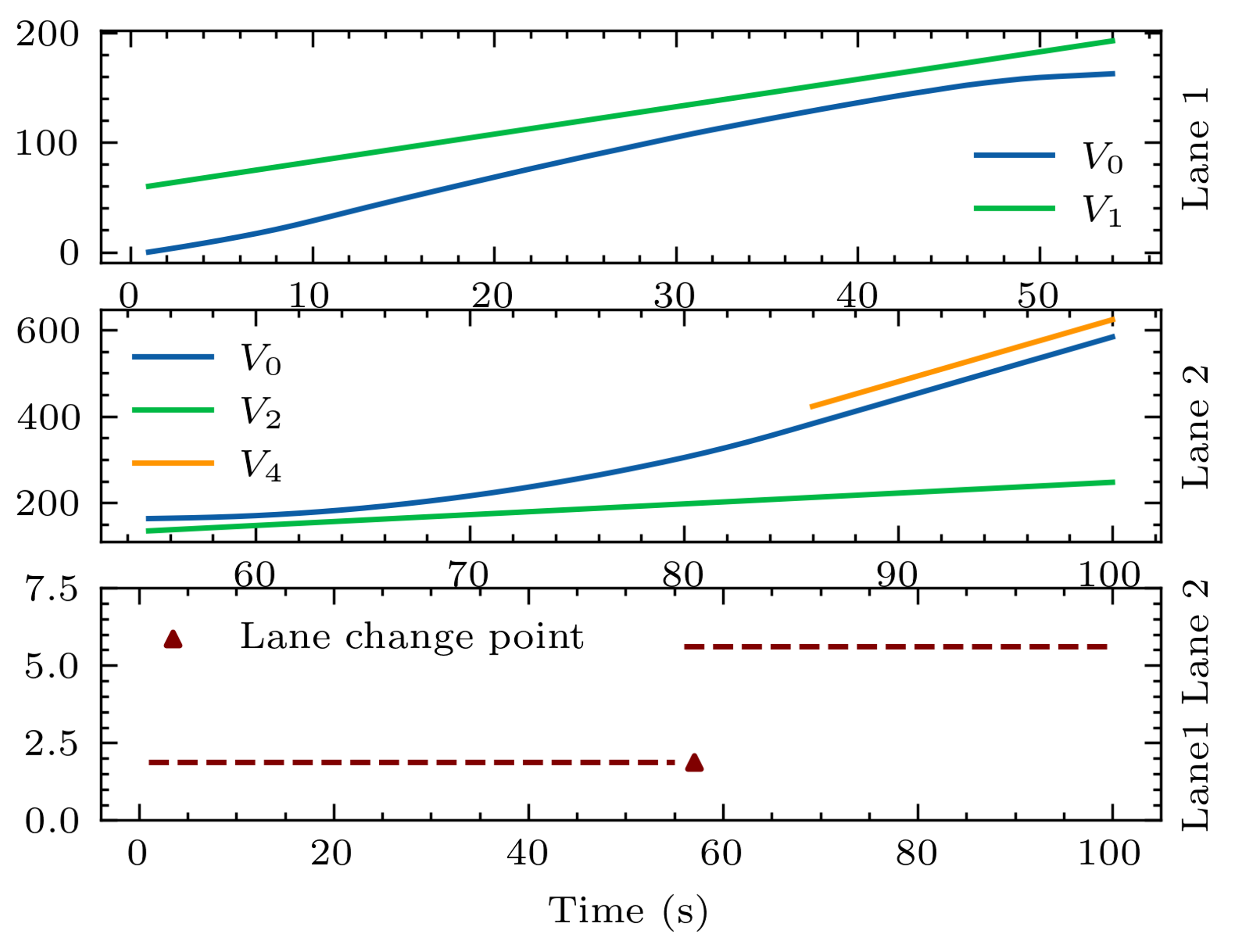}
            \caption{Scenario 4}
            \label{fig:6sub4}
        \end{subfigure}
    \end{minipage}
    
    \caption{Illustration for vehicles' spatial-temporal trajectory on the longitudinal direction and lane change time point}
    \label{fig:6}
\end{figure}

\begin{figure}[t] 
    \centering
    \begin{subfigure}[b]{0.8\textwidth}
        \includegraphics[width=\textwidth]{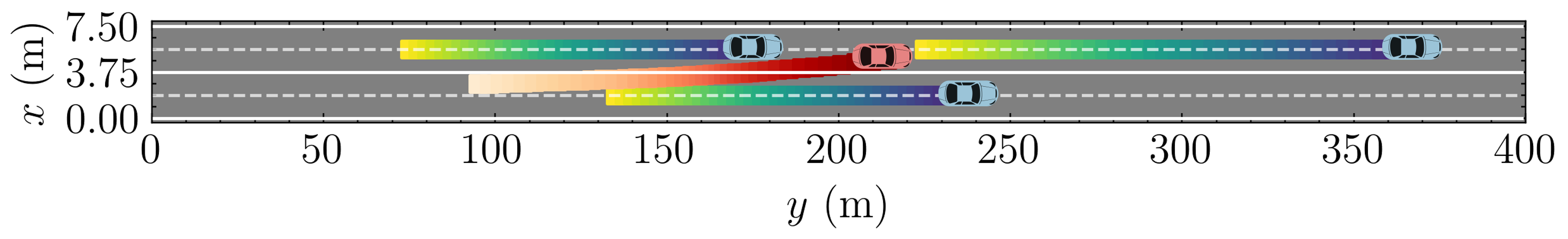}
        \caption{First lane change}
        \label{fig:7sub1}
    \end{subfigure}
    \hspace{0.05\textwidth} 
    \begin{subfigure}[b]{0.8\textwidth}
        \includegraphics[width=\textwidth]{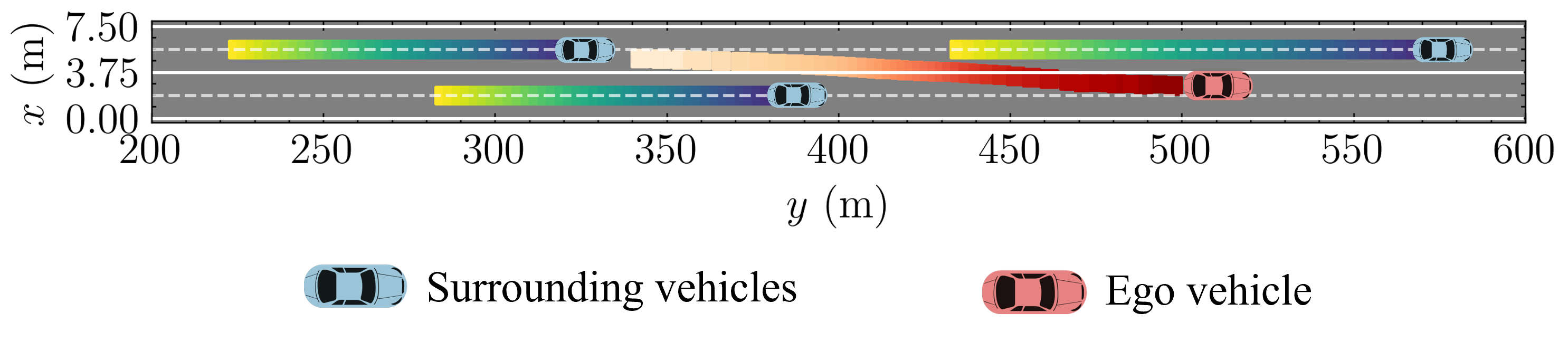}
        \caption{Second lane change}
        \label{fig:7sub2}
    \end{subfigure}
    \caption{Illustration for vehicles' movement in scenario 3(two lane changes)}
    \label{fig:7}
\end{figure}

Fig. \ref{fig:6} demonstrates the spatial-temporal trajectory of the vehicles in the longitudinal direction, which illustrates the effect of the integrated control of car-following and lane-changing behaviors. In our scenarios, the gap between $V_1$ and $V_2$ is acceptable for the ego vehicle to change from lane 1 to lane 2. However, this study initially positions the ego vehicle and $V_2$ at the same longitudinal location, which is similar to the example in the introduction part. Consequently, the ego vehicle cannot change lane at initial time step based on traditional methods such as MOBIL. However, the proposed integrated control framework can scheme for a lane change by adjusting the ego vehicle's longitudinal speed. As shown from Fig. \ref{fig:6}, at the beginning of the simulation, the ego vehicle does not immediately change to lane 2. Instead, it first accelerates to surpass $V_2$ in longitudinal direction and then utilizes the gap between $V_1$ and $V_2$. The successful lane change results from the algorithm learning the specific rewards associated with lane changing. In this reward structure, the ideal lane is identified based on the maximum potential speed of the ego vehicle on target lane. When the ego vehicle is not on the ideal lane, an extra reward for lane changing is conferred, motivating it to change lane. Through the integrated control of the car-following and lane-changing behavior, this lane change reward can be achieved by adjusting both longitudinal and lateral acceleration. Fig. \ref{fig:7} demonstrates the dynamics of vehicles in scenario 3 in an intuitive way, which can help understand spatial-temporal trajectory figures. The red rectangles highlight the lane change process. This simulation result demonstrates the problem proposed by the example in the introduction part is resolved.

The proposed P-DQN exhibits superior performance in simulating car-following behavior. In scenarios 2 and 4, vehicle $V_4$, preceding the ego vehicle $V_0$, is consistently maintained at a safe and constant distance following the completion of lane-changing maneuvers, as evidenced by Fig. \ref{fig:6}. Furthermore, $V_0$ successfully maintains a safe distance from preceding vehicles during lane changes. Specifically, in scenario 1 and in scenario 3, vehicles $V_1$ and $V_3$ serve as the leading vehicles to $V_0$, which is executing a lane change. The algorithm enables $V_0$ to mitigate collision risks with these leading vehicles while effectively exploiting the available gap for lane changing. This proficient behavior is attributed to the algorithm's capacity to learn from car-following rewards.

In addition to maintaining a safe distance from leading vehicles, $V_0$ also ensures an adequate distance from trailing vehicles during lane changes, as demonstrated with vehicle $V_2$ across all scenarios. Unlike the strategy for leading vehicles, the avoidance of $V_2$ is governed by learning from safety rewards, emphasizing the prevention of collisions with adjacent vehicles and road boundaries. To evaluate the safety learning objective, this study conducted a comprehensive assessment of the P-DQN's performance over 100 episodes for each scenario. The outcomes indicate an absence of collisions with surrounding vehicles and road boundaries, showcasing the P-DQN's capability to safely navigate through lane changes while maintaining stable algorithmic performance.

\begin{figure}[h]
    \centering
    \begin{minipage}{0.9\textwidth}
        \centering
        \begin{subfigure}[b]{0.45\textwidth}
            \includegraphics[width=\textwidth]{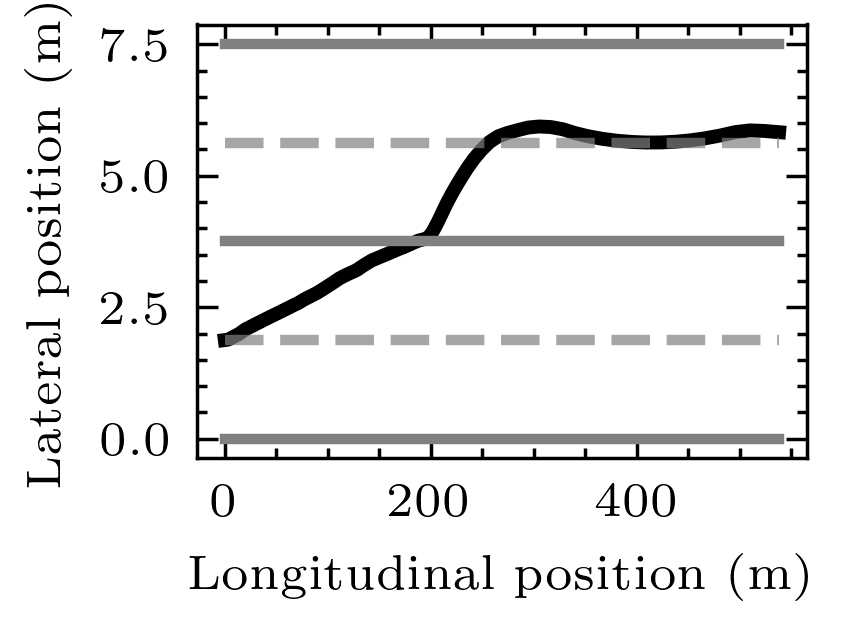}
            \caption{Scenario 1}
            \label{fig:8sub1}
        \end{subfigure}
        \hfill
        \begin{subfigure}[b]{0.45\textwidth}
            \includegraphics[width=\textwidth]{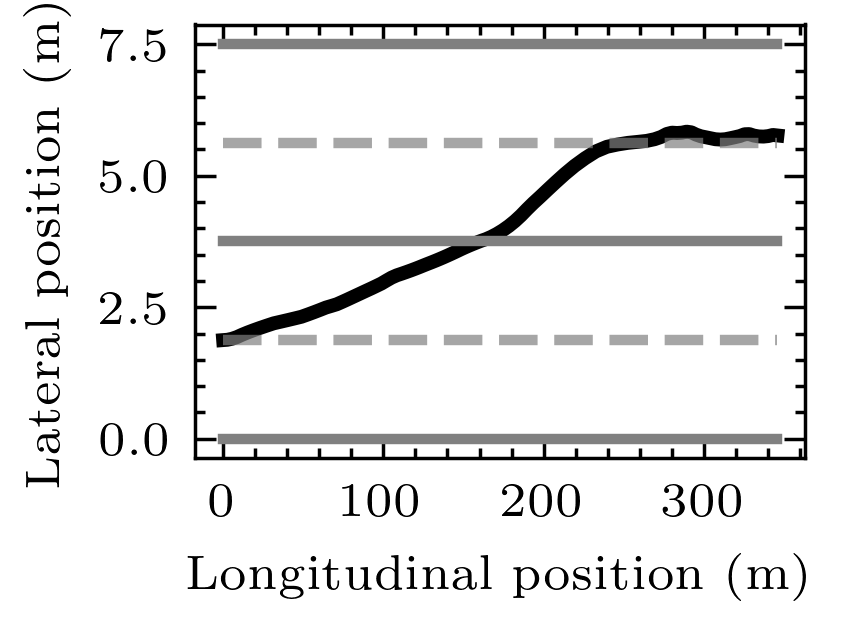}
            \caption{Scenario 2}
            \label{fig:8sub2}
        \end{subfigure}
    \end{minipage}
    
    \vspace{1em}
    
    \begin{minipage}{0.9\textwidth}
        \centering
        \begin{subfigure}[b]{0.45\textwidth}
            \includegraphics[width=\textwidth]{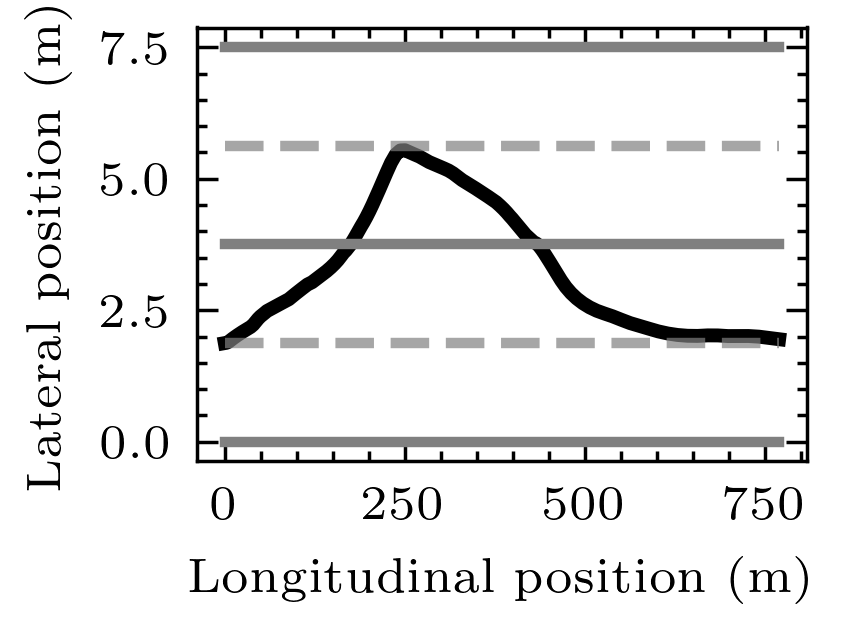}
            \caption{Scenario 3}
            \label{fig:8sub3}
        \end{subfigure}
        \hfill
        \begin{subfigure}[b]{0.45\textwidth}
            \includegraphics[width=\textwidth]{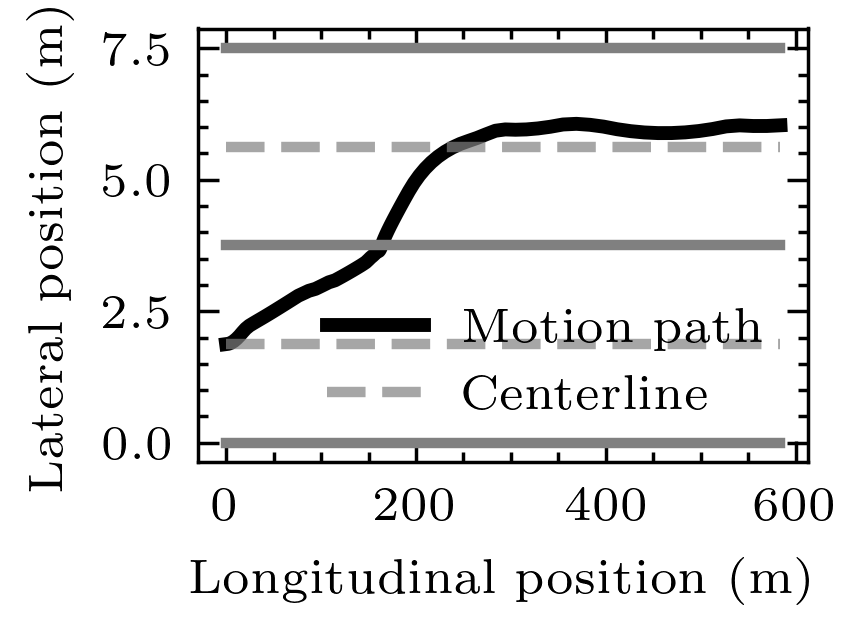}
            \caption{Scenario 4}
            \label{fig:8sub4}
        \end{subfigure}
    \end{minipage}
    
    \caption{Illustration for vehicles' lane change trajectory}
    \label{fig:8}
\end{figure}

The P-DQN model demonstrates enhanced efficacy in managing lane-keeping tasks, as evidenced by the data presented in Fig. \ref{fig:8}. The ego vehicle not only executes a seamless lane change but also maintains its trajectory along the centerline of the lane. This performance is attributed to the strategic integration of lane-keeping rewards within the model's learning framework. In scenarios devoid of lane-changing activities, the lane-keeping reward plays a pivotal role in modulating the vehicle's lateral acceleration to ensure stability and precision. Conversely, upon the initiation of a lane-change maneuver, both the lane-changing and lane-keeping rewards collaboratively influence the vehicle's lateral acceleration dynamics. Specifically, the lane-changing reward instigates the maneuver, while the lane-keeping reward facilitates the completion of this process, ensuring a smooth transition and sustained alignment within the lane. This dual-reward strategy underscores the model's ability to adaptively balance between maintaining lane integrity and executing necessary positional adjustments, thereby showcasing the P-DQN's superior performance. 

\subsubsection{Comparison with MOBIL+IDM}
\begin{figure}[H] 
    \centering
    \begin{subfigure}[b]{0.6\textwidth}
        \includegraphics[width=\textwidth]{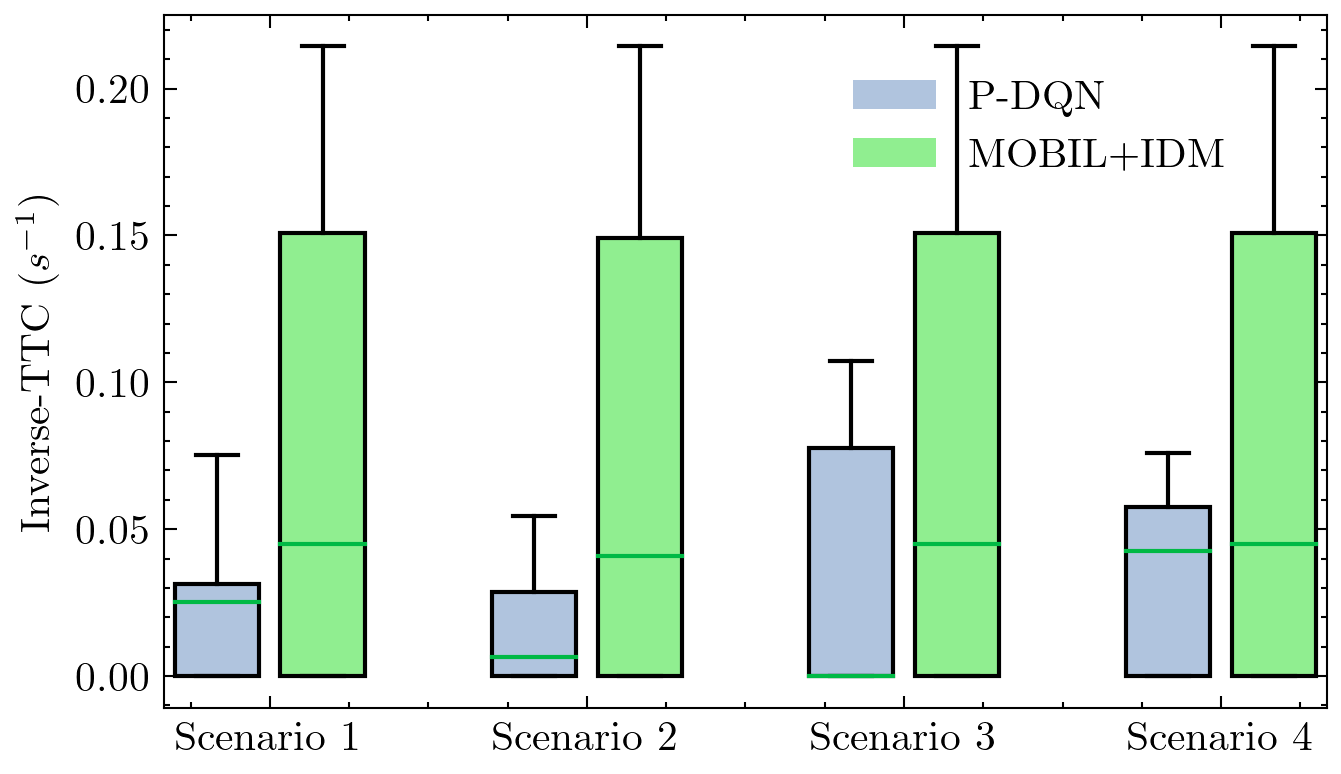}
        \caption{Inverse TTC distribution}
        \label{fig:9sub1}
    \end{subfigure}
    \hspace{0.05\textwidth} 
    \begin{subfigure}[b]{0.5\textwidth}
        \includegraphics[width=\textwidth]{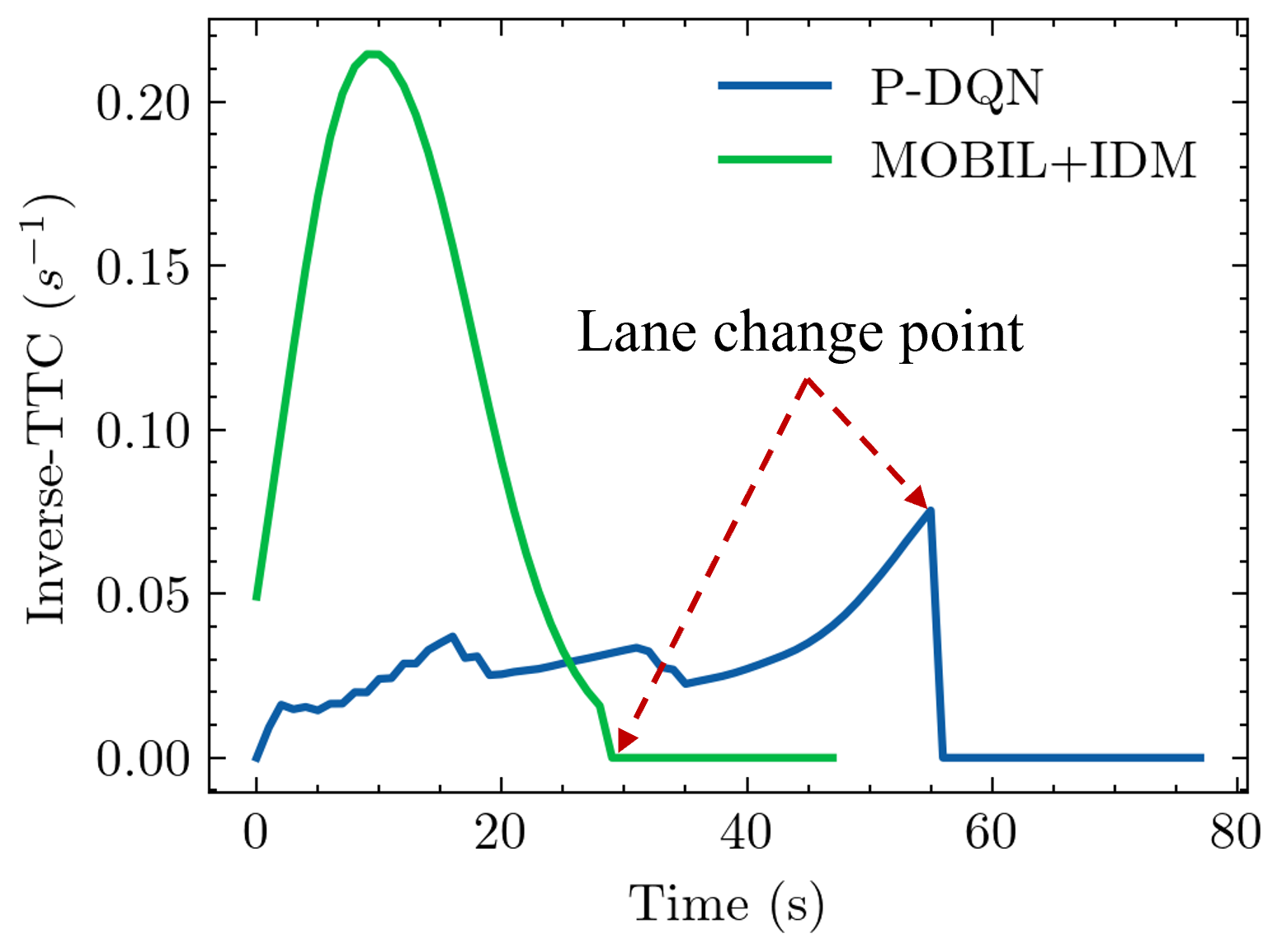}
        \caption{Inverse TTC}
        \label{fig:9sub2}
    \end{subfigure}
    \caption{Illustration for inverse-TTC's distribution (a) and change over time in scenario 1 (b) of MOBIL+IDM and P-DQN}
    \label{fig:9}
\end{figure}

The research utilizes the MOBIL+IDM framework as the baseline model, wherein MOBIL is deployed for executing lane change decisions, and IDM is applied for the regulation of longitudinal velocity. The ego vehicle is endowed with a predetermined lane change trajectory, which it adheres to throughout the lane change maneuver.  The baseline model undergoes evaluation across four distinct scenarios, employing two principal metrics, namely driving comfort and safety, to gauge its efficacy \citep{treiber2016mobil,kiefer2005developing}. Driving comfort is quantitatively defined, as delineated in Eq. \eqref{eq:21}, and assessed based on the magnitude of vehicle acceleration. Safety is gauged through the metric of inverse time-to-collision (inverse-TTC), as explicated in Eq. \eqref{eq:22}. Table \ref{table:4} delineates the comparative outcomes for both models, with inverse-TTC values indicative of the maximum values recorded in each episode. High inverse-TTC values signify an increased collision risk. Fig. \ref{fig:9sub1} provides a distribution of the inverse-TTC, evidencing the superior safety performance of P-DQN over the baseline model through consistently lower inverse-TTC values. Fig. \ref{fig:9sub2} further illustrates that P-DQN maintains low inverse-TTC values throughout individual episodes. The uniform low inverse-TTC values associated with P-DQN underscore its proficiency in ensuring a safe distance from surrounding vehicles, thereby enhancing overall vehicular safety.

\begin{equation}
C_i = \frac{1}{n} \sum_{t=0}^{n} \left( (a_{1,i}^t)^2 + (a_{2,i}^t)^2 \right)
\label{eq:21}
\end{equation}
\begin{equation}
T_i = \begin{cases} 
0, & \text{if } \Delta y_i^t \geq \delta_2 \text{ and } \Delta x_i^t \geq \delta_1 \\
\frac{\Delta v_{2,i}^t}{\Delta y_i^t}, & \text{if } \Delta y_i^t < \delta_2 \text{ and } \Delta x_i^t \geq \delta_1 \\
\frac{\Delta v_{1,i}^t}{\Delta x_i^t}, & \text{if } \Delta y_i^t \geq \delta_2 \text{ and } \Delta x_i^t < \delta_1 \\
\end{cases}
\label{eq:22}
\end{equation}

\begin{table}[tbp]
\centering
\caption{Comfort and safety performance of MOBIL+IDM and P-DQN}
\label{table:4}
\begin{tabular}{lllll}
\hline
Comfort & Scenario 1 & Scenario 2 & Scenario 3 & Scenario 4\\
\hline
MOBIL & 3.712 & 1.596 & 2.325 & 2.453 \\
P-DQN & 0.599 & 0.616 & 1.260 & 0.947 \\
\hline
Inverse-TTC & Scenario 1 & Scenario 2 & Scenario 3 & Scenario 4 \\
\hline
MOBIL & 0.214 & 0.214 & 0.214 & 0.214 \\
P-DQN & 0.075 & 0.054 & 0.107 & 0.076 \\
\hline
\end{tabular}
\end{table}

Where $a_{1,i}^t$ and $a_{2,i}^t$ denote the vehicle’s accelerations at two directions at step $t$, $n$ denotes the total steps in one episode. $\Delta x_i^t$ and $\Delta y_i^t$ are the relative distance between the ego vehicle and a surrounding vehicle on two directions, $\delta_{1}$ and $\delta_{2}$ are the threshold which are used to judge if there is a collision, $\Delta v_{1,i}^t$ and $\Delta v_{2,i}^t$ are the relative velocity at two directions.
\section{Conclusions}
This paper introduced a novel approach to integrate the control of lane-changing and car-following behaviors based on the P-DQN, which is adept at handling discrete-continuous hybrid action spaces. It differentiates between the discrete action of lane-change decision-making and the continuous actions of longitudinal and lateral accelerations for car-following and lane maintenance, respectively. By designing a hybrid action space that encompasses both discrete and continuous actions, the study effectively employs P-DQN to address the complexities involved in integrated control during lane-changing and car-following scenarios. For P-DQN development, this study designs the reward functions based on the learning objectives and the action space of the integrated control problem. The reward functions involve driving safety, car-following, lane-keeping, and lane-changing.

The proposed model is validated through four typical but representative numerical experiments, including car-following, lane-changing, lane keeping, and avoiding collisions. The P-DQN algorithm demonstrated convergence within reasonable training episodes for all scenarios. The problem incurred by the separate control of lane-changing and car-following is resolved by the proposed P-DQN. Simulation results revealed the P-DQN can accelerate or decelerate to perform a safe and efficient lane change due to the integrated control of lane-changing and car-following behaviors. Additionally, the P-DQN demonstrated the ability to reliably sustain lane-keeping and car-following behaviors when the ego vehicle was not engaged in active lane-changing. Furthermore, a comparative analysis with the MOBIL model highlighted the P-DQN’s superior performance in both driving comfort and safety.


\bibliographystyle{elsarticle-harv}
\bibliography{cas-refs}



\end{document}